\documentclass[nofootinbib,11pt,
      twoside,preprintnumbers,notitlepage]{revtex4-2}
\usepackage{epsfig,amsmath,amssymb}
\usepackage{tikz}
\usepackage{tikz-feynman}
\usepackage{color}
\usepackage{float}
\usepackage[colorlinks]{hyperref}
\usepackage{subfigure}
\usepackage[utf8]{inputenc}
\usepackage{slashed}
\usepackage{times}
\usepackage{epstopdf}
\usepackage{booktabs}
\usepackage{multirow}
\usepackage[normalem]{ulem}
\usepackage{natbib}
\usepackage[T1]{fontenc}
\usepackage{subfigure}
\usepackage{diagbox}
\usepackage{comment}

\hypersetup{
    linktocpage,
     colorlinks,
     citecolor=darkgreen,
     linkcolor= darkgreen,
     urlcolor=darkgreen
}
\definecolor{darkred}{rgb}{0.5,0.0,0.0}
\definecolor{darkblue}{rgb}{0.0,0.0,0.9}
\definecolor{darkerblue}{rgb}{0.0,0.0,0.5}
\definecolor{purple}{rgb}{0.5,0.0,0.5}
\definecolor{darkgreen}{rgb}{0.0,0.5,0.0}
\definecolor{black}{rgb}{0.0,0.0,0.0}
\definecolor{brown}{rgb}{0.6,0.4,0.2}
\definecolor{newpurple}{rgb}{0.65, 0.38, 0.61}
\definecolor{newyellow}{rgb}{0.9718, 0.6093, 0.0759}
\definecolor{amber}{rgb}{1.0, 0.75, 0.0}
\definecolor{newblue}{rgb}{0.4, 0.52, 0.85}
\definecolor{newred}{rgb}{0.8524, 0.2595, 0.3294}
\definecolor{newgreen}{rgb}{0.2, 0.8, 0.2}
\definecolor{SMgreen}{rgb}{0.56, 0.69, 0.19}
\definecolor{neworange}{rgb}{0.94, 0.462, 0.162}

\definecolor{BrickRed}{rgb}{0.9,0.1,0}

\newcommand{\bea}{\begin{eqnarray}}
\newcommand{\eea}{\end{eqnarray}}
\newcommand{\beq}{\begin{equation}}
\newcommand{\eeq}{\end{equation}}
\newcommand{\ec}{\end{center}}
\newcommand{\bc}{\begin{center}}

\begin{document}

\preprint{ULB-TH/24-12}

\title{On new physics off the Z peak in $H \rightarrow \ell^+\ell^- \gamma$}

\author{Aliaksei Kachanovich$^1$, Jean Kimus$^1$, Steven Lowette$^2$ and Michel H.G. Tytgat$^1$} 

\date{\today}

\affiliation{$^1$Service de Physique Th\'eorique, CP225
Universite Libre de Bruxelles
Boulevard du Triomphe (Campus de la Plaine)
1050 Bruxelles
Belgium $^2$ Interuniversity Institute for High Energies (IIHE), Vrije Universiteit Brussel, Pleinlaan 2, 1050 Brussels, Belgium}  

\email{aliaksei.kachanovich@ulb.be, jean.kimus@ulb.be, steven.lowette@vub.be, michel.tytgat@ulb.be}

\begin{abstract}
Motivated by a small but intriguing excess observed in the decay mode \( H\rightarrow \ell^+\ell^- \gamma \) reported by both the ATLAS and CMS collaborations, we explore the possibility that new physics contributes directly to the effective \( H \ell \overline{\ell} \gamma \) coupling rather than modifying the \( Z \) peak. Concretely, we consider a dimension-8 operator that could arise from new particles via box diagrams. Such non-resonant contribution may provide an alternative origin for current or future excesses. We examine how experimental cuts may distinguish between possible modifications of the \( Z \) peak and non-resonant contributions. 
The currently measured excess requires that the new physics scale is relatively low (\( \Lambda_R \sim v \)). However, we show that it may remain within current experimental bounds. In particular, we illustrate this using a simplified model, motivated by the dark matter problem, and discuss its other experimental constraints.
\end{abstract}

\maketitle

\section{Introduction}

\label{sec:intro}

The Standard Model (SM) of fundamental interactions is a well-established theory but it remains incomplete. Of particular interest is the scalar sector at the origin of electroweak symmetry breaking and, in particular, the properties of the Brout-Englert-Higgs boson (Higgs, or \( H \) for short). New physics beyond the SM could affect its rare decay modes, for instance \( H \rightarrow Z \gamma \) \cite{Cahn:1978nz,Bergstrom:1985hp,Spira:1991tj}. First evidence for the rare decay \( H \rightarrow  Z\gamma \) has been reported recently by the ATLAS and CMS collaborations \cite{ATLAS:2020qcv,CMS:2022ahq}. The combined results from both experiments indicate a branching fraction \( \text{Br}_{\rm obs} = (3.4 \pm 1.1) \times 10^{-3} \) in this channel, which is \( 2.2 \pm 0.7 \) times higher than the SM prediction \cite{ATLAS:2023yqk}. Although this discrepancy corresponds to only \( 1.9 \) standard deviations, the fact that both experiments report a slight excess is intriguing. This has motivated several studies proposing possible explanations, including new physics \cite{Barducci:2023zml,Boto:2023bpg,Das:2024tfe,Cheung:2024kml,Hu:2024slu,Hernandez-Juarez:2024pty,Zhang:2024yqw,Arhrib:2024wjj,Israr:2024ubp,Mantzaropoulos:2024vpe,Sang:2024vqk}.

Experimentally, the process \( H \to Z \gamma \) is reconstructed by measuring dileptons, \( Z^{(\ast)} \rightarrow \ell^+ \ell^- \) with \( \ell=e, \mu \), along with a photon in the final state. The background for \( H \to Z^{(\ast)} \gamma \to \ell^+ \ell^- \gamma \) includes contributions from \( H \to \gamma^\ast \gamma \), box diagrams with direct contributions to \( H \rightarrow \ell^+\ell^- \gamma \), and Bremsstrahlung from the tree-level \( H \) decay into muon pairs, \( H \to \mu^+ \mu^- \gamma \) \cite{Kachanovich:2020xyg,Kachanovich:2021pvx,Corbett:2021iob,Chen:2021ibm,Ahmed:2023vyl,Aakvaag:2023xhy,Hue:2023tdz,VanOn:2021myp,Sun:2013cba,Phan:2021xwc,Phan:2021ovj,Kachanovich:2024vpt,Abbasabadi:1996ze,Chen:2012ju,Dicus:2013ycd,Passarino:2013nka,Han:2017yhy}. Currently, the \( Z \) contribution is reconstructed by requiring a dilepton invariant mass above \( 50 \, \text{GeV} \) \cite{ATLAS:2020qcv,CMS:2022ahq,ATLAS:2023yqk}, which significantly reduces the \( H \to \gamma^\ast \gamma \) background  but does not remove it completely, as suggested in Figure \ref{fig:rate_SM}, which depicts the partial decay rate \({d\Gamma}/{dm_{\ell\ell}} \) as a function of the dilepton invariant mass \( m_{\ell\ell} \).

It has been shown that NLO QCD \cite{Buccioni:2023qnt} and EW \cite{Chen:2024vyn} contributions to this channel are negligible. Barring the possibility of a statistical fluctuation, the physical explanations for the observed excess can be broadly classified into two categories. One involves modifications to the decay rate \( H \rightarrow Z \gamma \), due to mixing with new scalars \cite{Das:2024tfe}. A key difficulty for such framework is modifying the \( HZ\gamma \) effective coupling while leaving the well-measured \( H \gamma^{{\ast}} \gamma \) coupling unaffected. To overcome this issue, one possibility is to consider that the Higgs  may decay into a \( Z \) and a light scalar particle, \( H \rightarrow Z a \). If \( m_a \sim \) GeV or less, the decay \( a \rightarrow \gamma \gamma \) — an expected standard feature of axion-like particles — produces very collimated photons, which may be mistaken for a single photon in the final state \cite{Cheung:2024kml, Alonso-Alvarez:2023wni}.

Although significant new physics in this channel is unlikely, it is important to consider all possibilities. In this paper, we explore whether new physics may contribute directly to the \( H \ell^+ \ell^- \gamma \) effective coupling. This would constitute a new background to the measurement of \( H \rightarrow Z \gamma \), a possibility that is admittedly ad hoc but which, as we will show, is not excluded by the current experimental situation.  We begin our analysis in general terms, using a dimension-8 effective operator, see equation \eqref{eq:effOP}.  We show to which extent experimental cuts can allow to distinguish between the two scenarios. Because \( H \rightarrow \ell^+ \ell^- \gamma \) is a loop suppressed process,  to compete the new physics scale  must be close to the EW scale, \( \Lambda_R \sim v = 246 \, \text{GeV} \). As the validity of the effective operator is questionable at this energy scale, we consider a simple UV model inspired by the dark matter problem. We discuss its signatures, constraints and other caveats before drawing some conclusions. Two appendices are dedicated to other aspects of the UV model.

\section{Standard Model contributions to $H\rightarrow \ell^+\ell^- \gamma$}

\label{sec:SM}

\begin{figure}[t]
	\begin{center}
		\subfigure[t][]{\includegraphics[width=0.23\textwidth]{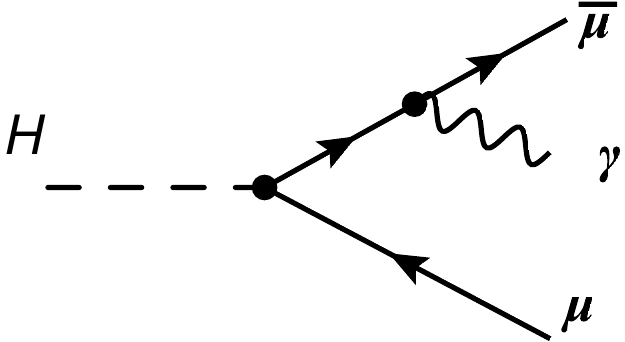}}
		\hspace{.6cm}
  	\subfigure[t][]{\includegraphics[width=0.23\textwidth]{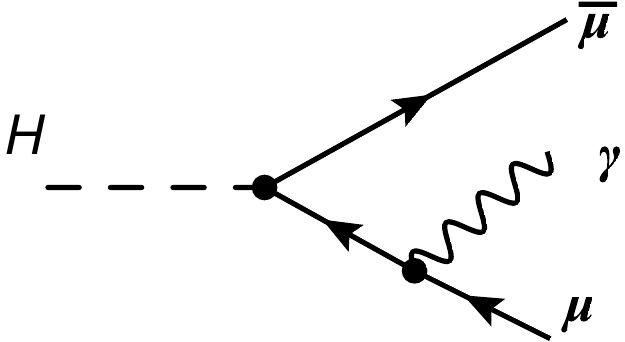}}
         \end{center}
	\caption{Bremsstrahlung from final state muon pairs. The contribution from electron pairs is negligible due to the smallness of the electron Yukawa coupling.}
	\label{fig:tree}
\end{figure}

\begin{figure}[t]
	\begin{center}
		\subfigure[t][]{\includegraphics[width=0.23\textwidth]{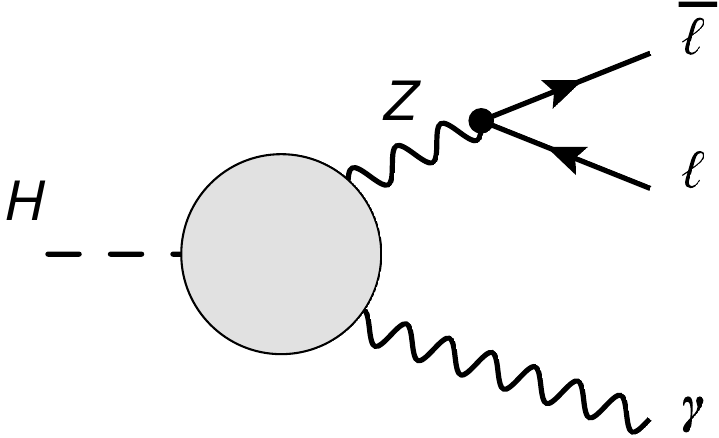}}
		\hspace{.6cm}
  	\subfigure[t][]{\includegraphics[width=0.23\textwidth]{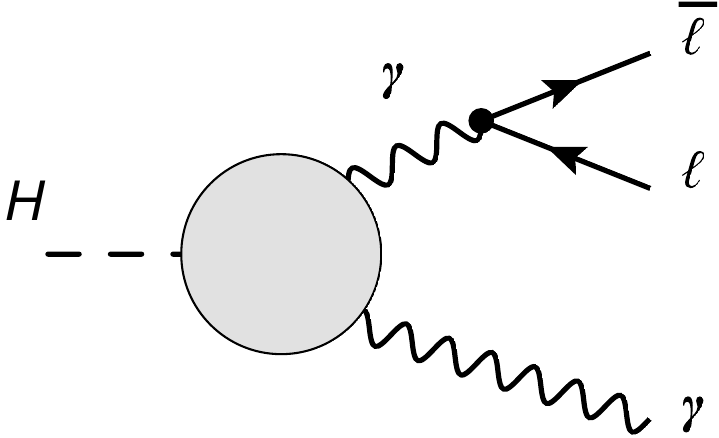}}
   		\hspace{.6cm}
   		\subfigure[t][]{\includegraphics[width=0.23\textwidth]{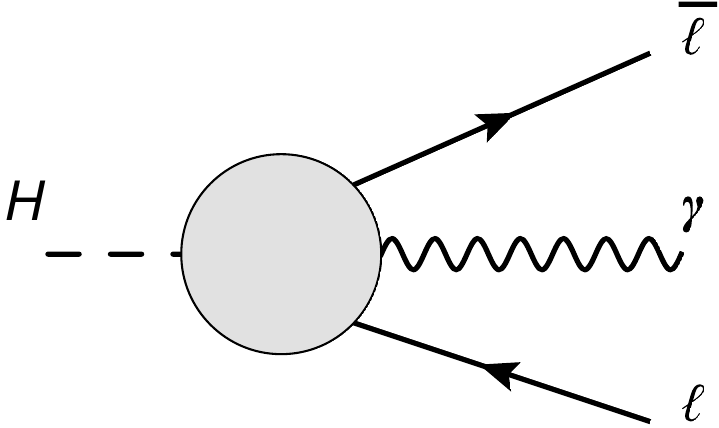}}
         \end{center}
	\caption{Loop amplitudes (schematic) contributing to $H \rightarrow \ell^+\ell^- \gamma$. In the SM, (a) and (b) arise  at one-loop through triangular diagrams and (c) from box diagrams. Both (b) and (c) are non-resonant. }
	\label{fig:LoopSM}
\end{figure}

 In the SM, \( H \to \ell^+ \ell^- \gamma \) arises from four distinct sub-processes. First, through Bremsstrahlung from muon pairs, see Figure \ref{fig:tree}. Due to the smallness of the electron Yukawa coupling,  this process constitutes a negligible contribution for electrons pairs.  Next, at loop level, there are two distinct triangular diagrams, leading respectively to (a) in Figure \ref{fig:LoopSM}) $H \to Z^{(\ast)} \gamma$ and (b) $H \to \gamma^\ast \gamma$. Last, there are box diagrams that lead to  direct $H \ell\bar \ell \gamma$ coupling (c). 
 Up to gauge-dependent parts, the separation between these three kinds of subprocesses is rather well defined. The process of interest is of course the resonant decay through the $Z$ boson.  
 %In particular, the $Z$ coupling gives rise to a resonant decay. We  refer to non-resonant new physics, we will specifically mean contributions  as (c) in fig.\ref{fig:LoopSM}. 
 
 The total one-loop amplitude has the following structure
\begin{eqnarray}
    \mathcal{M}_{\text{SM,loop}} &=& \nonumber \left[q_{\mu} p_{1} \cdot \varepsilon^{*}(q) - \varepsilon^{*}_\mu(q) \, q\cdot p_{1}\right] \bar{u}(p_{2}) \big( A_{1} \gamma^{\mu} P_{R} + B_{1} \gamma^{\mu} P_{L} \big) v(p_{1}) \\
    &+& \left[q_{\mu} p_{2} \cdot \varepsilon^{*}(q) - \varepsilon^{*}_\mu(q) \, q\cdot p_{2}\right] \bar{u}(p_{2})\big( A_{2} \gamma^{\mu} P_{R} + B_{2} \gamma^{\mu} P_{L} \big) v(p_{1}) \,,
    \label{eq:SMloop}
\end{eqnarray}
 using standard textbook notations \cite{Peskin:1995ev}. The four-momenta of the photon, lepton and anti-lepton are denoted by $q$, $p_{1}$, and $p_{2}$, respectively. This expression makes it manifest that  the amplitude satisfies the Ward identity and so vanishes for $\epsilon^\ast(q) \rightarrow q$.  
The form factors  $A_{1,2}$ and $B_{1,2}$ depend on Mandelstam variables that are defined as follows.  The squared invariant masses are represented by  $s=(p_1 + p_2)^2$, $ t=(p_1 + q)^2$, and $ u=(p_2 + q)^2$, which satisfy the relation $ s+t+u = m_H^2 + 2 m_\ell^2  \approx m_H^2 $, with  $m_{H}$  the $H$ boson mass and $m_\ell$ that of the leptons. The coefficients $A_2$ and $B_2$ are obtained from $A_1$ and $B_1$ by interchanging $t$ and $u$. Explicit expressions for the coefficients $A_{1,2}$ and $B_{1,2}$ can be found in \cite{Kachanovich:2020xyg}.

%\subsection{Numerical results}
\begin{figure}[]
{\includegraphics[width=0.9\textwidth]{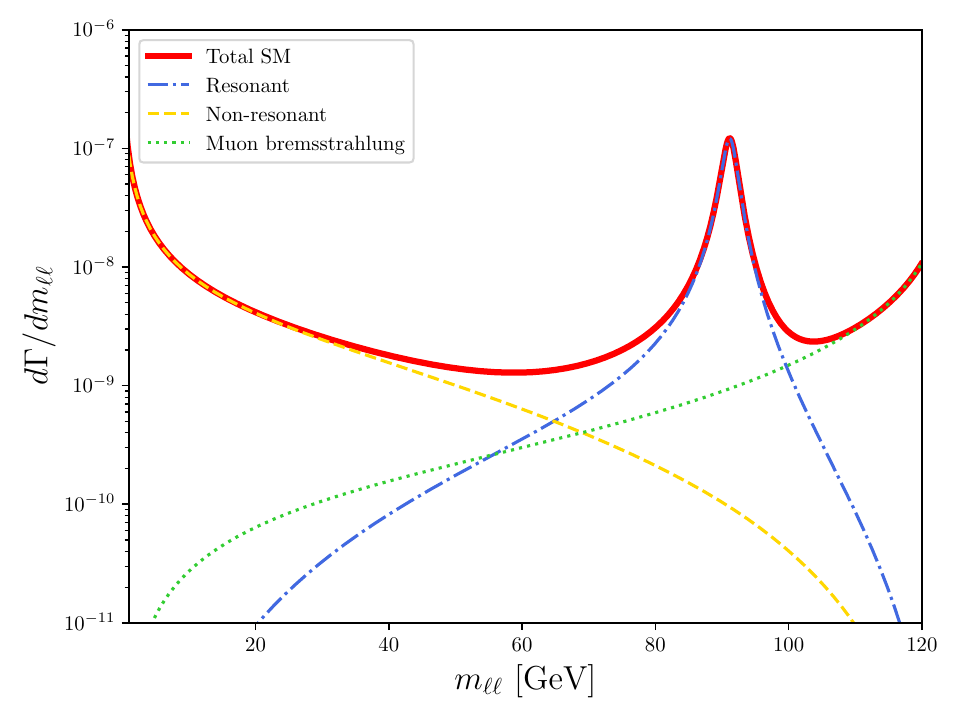}}
	\caption{Differential decay rate $d\Gamma/dm_{\rm \ell\ell}$ for $H \rightarrow \ell^+ \ell^- \gamma$ as function of the dilepton inviariant mass, summed (solid red line) for electron and muon pairs. The blue dot-dashed curve corresponds to the resonant $Z$ contribution, see the text. The dotted green curve corresponds to  Bremsstrahlung from muon pairs. The dashed yellow curve corresponds to the non-resonant contributions, which include $H \rightarrow \gamma^\ast \gamma$ and the contribution from box diagrams. We emphasize that this curve is theoretical as it assumes no cuts on the energy of the leptons and of the photon. }
	\label{fig:rate_SM}
\end{figure}

The SM theoretical prediction for the differential decay rate \( {d\Gamma}/{dm_{\ell\ell}} \) as a function of the dilepton invariant mass is shown in Figure \ref{fig:rate_SM}. 
This figure is theoretical as it assumes that no cuts are set on the energy of the leptons and of the photon. The total contribution, which is obtained by summing the electron pair and the muon pair channels, is represented by the solid (red) curve. The dotted green curve is Bremsstrahlung from muon pairs, which peaks at large \( m_{\rm \ell\ell} \sim m_H \), corresponding to emission of soft photons. The resonant production from the \( Z \)-boson, modeled with a Breit-Wigner distribution, is shown by the dot-dashed (blue) curve. The long-dashed yellow curve corresponds to dilepton production from an off-shell photon and direct dilepton production (box diagrams), which are collectively labeled as non-resonant.

While the low \( m_{\rm \ell\ell} \) part of the non-resonant curve corresponds clearly to the contribution from \( H \rightarrow \gamma^\ast \gamma \), the distinction with contributions from the box diagrams is less clear at higher values of \( m_{\rm \ell\ell} \). Moreover, the different amplitudes involve gauge-dependent terms that cancel with each others. Therefore, a separation between resonant and non-resonant contributions requires some explanation. Here, we use the procedure explained in greater details in \cite{Kachanovich:2021pvx}. Concretely, we divide the amplitudes into two subsets which, in terms of the form factors defined in eq. \eqref{eq:SMloop}, correspond to  
\begin{eqnarray}
    A_{1(2)} (s,t) =  A_{1(2)}^{res}(s) + A_{1(2)}^{nr}(s,t) \,, 
\end{eqnarray}
with 
\begin{eqnarray}
     A_{1(2)}^{res}(s) \equiv \frac{\alpha (m_{Z}^2)}{s - m_{Z}^2 + i m_{Z} \Gamma_Z} \qquad \mbox{\rm and} \qquad A_{1(2)}^{nr}(s,t)\equiv \tilde{A}_{1(2)} (s, t) + \frac{\alpha(s) - \alpha (m_{Z}^2)}{s - m_{Z}^2 + i m_{Z} \Gamma_Z}\,, 
\end{eqnarray}
here, $\alpha (s)$ is a part of coefficients $A$, which is the same for $A_1$ and $A_2$. 
Analogously, we separate into resonant and non-resonant parts the form factor $B$. The original {form factor} can then be rewritten as 
\begin{eqnarray}
    A_{1(2)}(s,t) = \tilde{A}_{1(2)}(s,t) + \frac{\alpha (s)}{s - m_{Z}^2 + i m_{Z} \Gamma_Z}\,.
\end{eqnarray}
with a similar expression for $B_{1(2)}(s,t)$.
We emphasize that the resonant contribution depends only on the dilepton squared invariant mass and so on the variable $s$. 

Doing so, we can interpret the resonant contribution as the gauge independent part of the $H \to Z \gamma \rightarrow  \ell^+ \ell^- \gamma$ subprocess, which is shown by the dot-dashed blue curve in Figure \ref{fig:rate_SM}. %The gauge dependence cancels out only after summation of contribution of all diagrams.  
We use this separation between resonant and non-resonant contributions in the next section.

\section{New physics contributions to $H\rightarrow \ell^+\ell^- \gamma$}

\subsection{$Z$ peak new physics}
\label{sec:resc}
\begin{figure}[h]
{\includegraphics[width=0.9\textwidth]{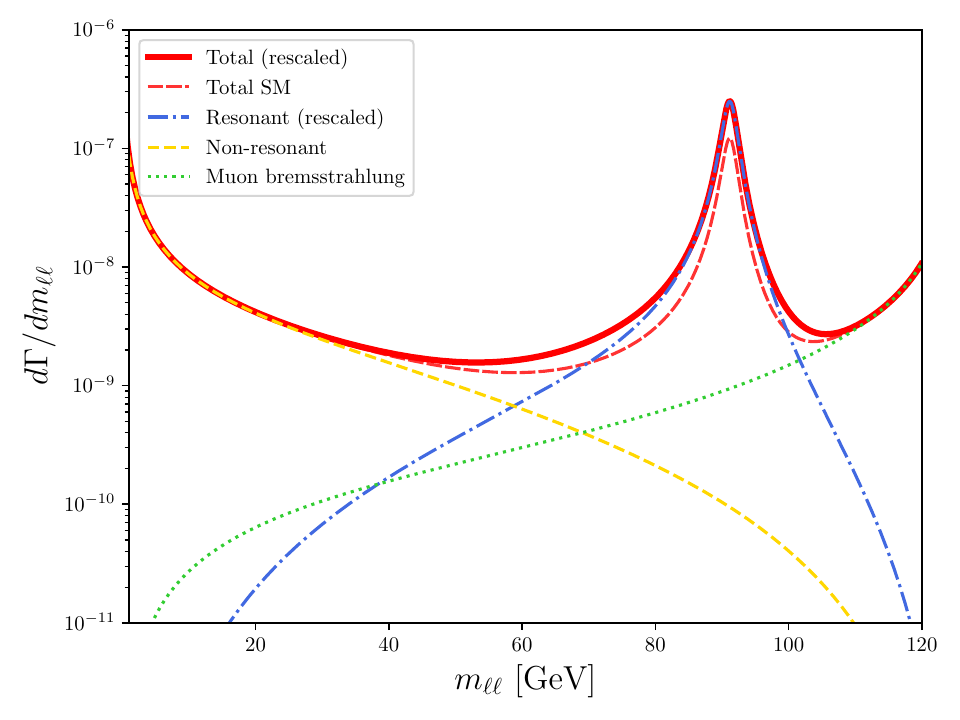}}
	\caption{Same as Figure \ref{fig:rate_SM}, together with a decay rate with the $Z$-peak contribution rescaled   to match the measured excess of events  (solid red curve) compared with the SM prediction (short-dashed red curve). } 
	\label{fig:rate_NP1}
\end{figure}

The  ATLAS and CMS collaborations aim to measure the effective coupling \( H Z \gamma \). To achieve this, they apply kinematic cuts to reduce background contributions and interpret the observed signal as originating from the  \( Z \) peak. Several Standard Model (SM) processes contribute to the background. We focus specifically on the background arising from Bremsstrahlung off muon pairs and the non-resonant subprocesses introduced in the previous section. As far as we know, the collaborations use a narrow-width approximation to extract the contribution from the \( Z \)-peak. They then report a deviation in the signal strength \( \mu \equiv {\text{Br}_{\rm obs}}/{\text{Br}_{\rm SM}} = 2.2 \pm 0.7 \), which is thus interpreted to be entirely due to the $Z$ peak, i.e. $\mu = \Gamma_{H \rightarrow Z\gamma}\vert_{\rm obs}/\Gamma_{H \rightarrow Z\gamma}\vert_{\rm SM}$. 

{Here instead, we put the focus  on the  events observed,  $\mu \equiv \Gamma_{H \rightarrow l^+l^-\gamma}\vert_{\rm obs}/\Gamma_{H \rightarrow l^+l^-\gamma}\vert_{\rm SM}$ given the experimental cuts. Then, using the separation between resonant and non-resonant contributions, as defined in the previous section, we leave the non-resonant and Bremsstrahlung subprocesses unchanged but rescale the resonant part by a factor equal to $2.11$, a value adjusted so as to obtain the central value of the measured events.  The slight difference between our rescaling factor and the value $2.2$ reported by the experiments, which is insignificant given the current uncertainties,  is due both to  the wings of the rescaled $Z$-boson resonance, as well as the non-resonant and bremsstrahlung contributions. The resulting rescaled differential decay rate \({d\Gamma}_{\rm resc}/{dm_{\rm \ell\ell}}\) is shown as the solid (red) line shown in Figure \ref{fig:rate_NP1}, together for comparison with the SM prediction (dashed red line)  as was shown in Figure \ref{fig:rate_SM}.}

In Table \ref{tab:sample}, we report our values for the ratio \( {\text{Br}_{\rm resc}}/{\text{Br}_{\rm SM}} \) for various cut choices, in particular on the invariant mass $m_{\rm \ell\ell}$. The first two lines ($1$ and $2$) are theoretical as no cuts are put on the energy of the final states particles. They thus correspond to numbers obtained by integrating curves in Figure \ref{fig:rate_NP1}. Line $4$, in bold, corresponds to the cuts applied by the CMS collaboration, in particular \(  50 \, \text{GeV} \leq m_{\rm \ell\ell}\leq  125 \, \text{GeV} \). The cuts on the energy of the final-state particles are  $E_1 \geq 7$ GeV, $E_2 \geq 25$ GeV and $E_\gamma \geq 15$ GeV (the ATLAS collaboration is using $E_\gamma \geq 10$ GeV, which essentially leads to the same result). The column $\Gamma_{\rm tot}^{\rm SM}$ corresponds to the total (resonant, non-resonant and Bremsstrahlung processes included) $H$ decay rate {\em in the SM}, given the cuts and calculated following the results  recapitulated  in Section \ref{sec:SM}. The column $\Gamma_{\rm tree}^{\rm SM}$ is the  contribution  from Bremsstrahlung and provides, for illustration,  the impact of this specific background. The ratio we obtained from the rescaled  $Z$ resonance is  \( {\text{Br}_{\rm resc}}/{\text{Br}_{\rm SM}} = 2.06\), a value that, given the current uncertainties, is  consistent with the value $\mu = 2.2$ reported by the collaborations.\footnote{Lest there is risk of confusion, we emphasize that the ratio $\frac{Br_{\rm resc}}{Br_{\rm SM}}$ presented in Table \ref{tab:sample} corresponds to the contribution of the rescaled $Z$-resonant part relative to the {\em same} quantity as predicted by the SM. Thus, this ratio does not take into account the non-resonant and bremsstrahlung contributions. This explains the difference between the value of the rescaling factor $2.11$ quoted above and the value $2.06$ given in Table \ref{tab:sample}.} The ${\cal O}(5 \%)$  difference is irrelevant given the current uncertainties but can be traced to be due to the Bremsstrahlung and non-resonant backgrounds (see Table \ref{tab:binsCMS} in Section \ref{sec:effOP}). Lines $3$ and $5-7$ correspond to different choices of cuts on the dilepton invariant mass. All other things kept constant, narrowing down around the $Z$ peak has a little impact on the signal strength, which remains $\mu \approx 2.1$, except in the non-realistic case of no cuts (line $1$). The last two columns of the table are discussed in sections \ref{sec:effOP} and \ref{sec:UV} below.

\begin{table}[ht!]
\centering
\begin{tabular}{|c|c|c|c|c|c|c|c|c|}
\hline
\,\#\,& {Cuts} & $m_{\ell \ell}^{min}$[GeV] & $m_{\ell \ell}^{max}$[GeV]  & $\Gamma_{\rm tot}^{\rm SM}$ [keV] & $\Gamma^{\rm SM}_{\rm tree}$[keV] & {$\frac{Br_{\rm resc}}{Br_{\rm SM}}$} & {$\frac{Br_{\rm EFT}}{Br_{\rm SM}}$} & {$\frac{Br_{\rm UV}}{Br_{\rm SM}}$}\\
\hline
1 & \, None \, & 50 & 125  & 0.768 & 0.287 & 1.67 & 1.86 & 2.07\\
\hline
2 & None & 50 & 100  & 0.504 & 0.028 & 2.01 & 2.21 & 2.57\\
\hline
3 & CMS & 40 & 125  & 0.455 & 0.011 & 2.04 & 2.10 & 2.13 \\
\hline\hline
\textbf{4} & \textbf{CMS} & \textbf{50} & \textbf{125} & \textbf{0.451} & \textbf{0.011}  & \textbf{2.06} & \textbf{2.06} & \textbf{2.06} \\
\hline\hline
5 & CMS & 70 & 125   & 0.440 & 0.011  & 2.07 & 1.80 & 1.71\\
\hline
6 & CMS & 70 & 100   & 0.432 & 0.006  & 2.08 & 1.74 & 1.68 \\
\hline
7 & CMS & 80 & 100   & 0.416 & 0.005  & 2.09 & 1.48 & 1.39\\
\hline
\end{tabular}
\caption{Signal strengths for various cuts choices.  {The first two lines assumes no cuts on the kinematical parameters, except $m_{\rm \ell\ell}$ and thus corresponds to values obtained by integrating under the curves given in Figures \ref{fig:rate_SM},\ref{fig:rate_NP1} and \ref{fig:rate_NP1}. Putting cuts suppresses the decay rate, in particular toward low values of $m_{\rm \ell\ell}$. In lines 3 to 7, we report numbers using the cuts of CMS : $E_\gamma \geq 15$ GeV, $E_1 \geq 7$ GeV, $E_2 \geq 25$GeV and $t_{min}, u_{min} \geq (0.1 m_H)^2$}. The  column $Br_{\rm resc}/Br_{\rm SM}$ corresponds to the expected signal strength from the  $Z$ peak over that of the SM assuming that the effective coupling of the $H Z \gamma$ coupling is rescaled by a factor of {$\sqrt{2.1}$}, see text for explanations. The  column $Br_{\rm EFT}/Br_{\rm SM}$ corresponds to the signal strength using the effective operator \eqref{eq:effOP} of section \ref{sec:effOP}. The column $Br_{\rm UV}/Br_{\rm SM}$ corresponds to the case of the UV model discussed in section \ref{sec:UV} using $m_S = m_F = 62.5$ GeV.}
\label{tab:sample}
\end{table}

\subsection{Non-resonant new physics}
\label{sec:effOP}

We now consider other possible sources of new physics in $H \rightarrow \ell^+\ell^- \gamma$,  but away from the $Z$ peak.  We begin with an effective operator description. There is a large body of literature on effective operator deviations from the Standard Model, starting with \cite{Grzadkowski:2010es}. Here we aim to be illustrative, not exhaustive and consider  a very specific dimension-8 operator,
\begin{equation}\label{eq:effOP}
   {\cal L}_{\rm eff} \supset {g'\over \Lambda_R^4} \vert \Phi  \vert^2 \partial_{\nu} (\bar \ell_{R} \gamma_{\mu} \ell_R) B^{\mu \nu} \,.
\end{equation}
Here, $\Phi$ is the SM Higgs doublet, {$\ell_R$} are $SU(2)$ singlets right-handed leptonic spinors, $B^{\mu\nu}$ is the  hypercharge field strength and $g' = e/\cos\theta_W$. The dimension $6$ part on which this operator is built, $\partial_{\nu} (\bar l_{R} \gamma_{\mu} l_R) B^{\mu \nu}$, vanishes for fields on-shell; the factor $\vert \Phi \vert^2$ is thus all important.  In the broken phase, $\vert \Phi\vert^2 \rightarrow v H + \ldots$, this operator contributes to the process $H \rightarrow \ell^+ \ell^- \gamma$.
\begin{figure}[]
{\includegraphics[width=0.9\textwidth]{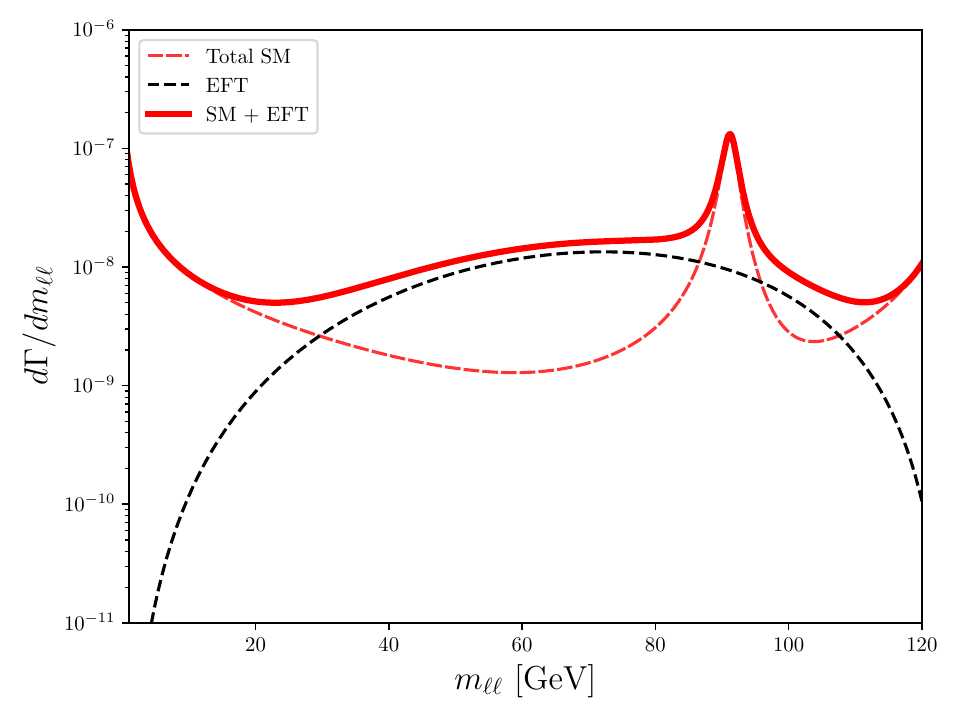}}
	\caption{Contribution of the effective operator \eqref{eq:effOP} to $H \rightarrow \ell^+\ell^- \gamma$. In this figure, no cuts are set on the kinematics of the outgoing particles, but the new physics scale has been set to $\Lambda_R = 260 \, \rm GeV$  so as to give the observed  excess of events compared to the SM prediction. }
	\label{fig:rateEffOp}
\end{figure}
We could have considered similar operators, involving left-handed leptons and the $SU(2)$ gauge field. The one we discuss is just a simple option (see \cite{Corbett:2021iob} for a discussion of dim 8 operators in the context of $H \rightarrow \ell^+ \ell^- \gamma$ decay). More importantly, we assume that the effective operator \eqref{eq:effOP} with coupling to electrons and muons has the same scale $\Lambda_R$, so as to avoid  lepton flavour violation in  $H \rightarrow \ell^+\ell^- \gamma$. This is clearly an important caveat of our  scenario but, doing so, the contribution of these effective operators to the  amplitude takes the form
\begin{equation}
 \Delta \mathcal{M} = {g' v \over \Lambda_R^4}\big[(q_{\mu} p_{1 \nu} - g_{\mu \nu} \, q\cdot p_{1}) - (q_{\mu} p_{2 \nu} - g_{\mu \nu} \, q\cdot p_{2})\big]\bar{u}(p_{2}) \gamma^{\mu} P_{R}  v(p_{1}) \varepsilon^{* \nu}(q)
\end{equation}
Comparing with eq.~\eqref{eq:SMloop}, we see that the contributions to the form factors are $\Delta B_1 = - \Delta B_2 = g' v/\Lambda_R^4$ and  $\Delta A_1 = \Delta A_2 = 0$. 

Given an appropriate value for $\Lambda_R$, the  impact of the effective operators on  \( d\Gamma/dm_{{\ell \ell}} \) is depicted in Figure \ref{fig:rateEffOp},  again assuming  no experimental cuts. The {dot-dashed black} bumpy line corresponds to the contribution of the effective operator only. The solid red line represents the total contribution, just obtained by adding the SM and the contribution of the effective operator, thus neglecting (for simplicity) interference effects. 
In the figure, we have adjusted the scale \( \Lambda_R \) so as to match the excess of events currently reported by ATLAS and CMS. Its value is  close to the EW scale, \( \Lambda_R = 260 \, \text{GeV} \sim v \). A low effective scale is  expected. The SM signal, while loop suppressed, is dominated by the $Z$ peak, the excess observed is important and we attempt to explain it using a dimension 8 effective operator. See also \cite{Boto:2023bpg} for general considerations on the scale of new physics required to explain the measured excess. 

Now, assuming the existence of such new physics and everything else being kept the same, it is  obvious that applying more stringent cuts, particularly on the dilepton invariant mass, should decrease the signal strength compared to the SM expectations. This is shown in the column  $Br_{\rm EFT}/Br_{\rm SM}$ of Table \ref{tab:sample}. Again, line $4$ corresponds to the current cuts used by CMS (or ATLAS), and as stated above, we have set  \( \Lambda_R \) so as to get the same excess of events as in the case of a rescaled $Z$ peak. More stringent cuts on $m_{\rm \ell\ell}$, in particular {those of line} $7$, lead to a strong suppression of the signal strength. Relaxing the cuts like in line $3$ gives a slightly (but  non-significant) larger signal strength. We emphasize again that like Figure \ref{fig:rate_SM}, no cuts are put on the energy of the outgoing particles in Figure \ref{fig:rateEffOp}. The hypothetical corresponding  signal strengths are given for illustration in lines $1$ and $2$. They are obtained by integrating under the curves shown in Figure \ref{fig:rateEffOp}.

Clearly, a large new physics contribution like the one we are considering is  an unlikely event. But, looking ahead, it may be good to have in mind that applying overly stringent cuts is perhaps not the optimal strategy when looking for new physics in $H$ decay. In Table \ref{tab:binsCMS}, we give the results we have obtained for the $H$ to $\ell^+\ell^- \gamma$ decay rate by integrating over bins of $10$ GeV, using the cuts of CMS. The first three lines correspond to the SM.  The last four lines correspond to the new physics scenarios considered in this paper. From these entries, one may reconstruct the signal strength reported in Table \ref{tab:sample}, as well as bin per bin. Table \ref{tab:bins} is given for reference, as it is the same as Table~\ref{tab:binsCMS} but involves no cuts; in principle it can be obtained by integrating the curves presented in Figures \ref{fig:rate_SM}, \ref{fig:rate_NP1} and \ref{fig:rate_NP2}, the latter being discussed in the next section. 

\begin{table}[ht!]
\centering
\begin{tabular}{|c|c|c|c|c|c|c|c|c|c|}
\hline
 & \, 40-50\,  & \, 50-60 \,   & \, 60-70 \,    & \, 70-80 \,   & \, 80-90 \,   & \, 90-100 \,   & \, 100-110 \,   & \, 110-125 \,  \\
\hline
 $\Gamma_{\rm SM}$  & 0.004 &  0.004 & 0.006  & 0.014 & 0.119 & 0.295 & 0.008 & 0.000  \\
 \hline
$\Gamma_{\rm tree}$   & 0.000 & 0.000 & 0.000  & 0.001 & 0.002 & 0.003 & 0.004 & 0.000  \\
 \hline
$\Gamma_{\rm NR}$   & 0.004 & 0.003 & 0.003  & 0.001 & 0.001 & 0.001 & 0.000  & 0.000  \\
\hline
\hline
$\Gamma_{\rm resc}$   & 0.001 & 0.002 & 0.006  & 0.023 & 0.242 & 0.621 & 0.008 & 0.000 \\
\hline
$\Gamma_{\rm EFT}$   & 0.024 & 0.046 & 0.078  & 0.119  & 0.118 &  0.079 & 0.035  & 0.000  \\
\hline
$\Gamma_{\rm UV62}$  & 0.040 &  0.067 & 0.098   & 0.129 & 0.103  & 0.053 & 0.017 & 0.000  \\
\hline
$\Gamma_{\rm UV100}$  & 0.027 & 0.048 & 0.080   & 0.121 & 0.117 & 0.074 & 0.031 & 0.000  \\
\hline
\end{tabular}
\caption{Binned decay rates (in keV), using CMS cuts,  for different intervals of dilepton invariant mass $m_{\ell \ell}$ (in GeV). The first 3 lines correspond to the SM. $\Gamma_{SM}$ is the total SM contribution, $\Gamma_{\rm tree}$ the contribution from Bremsstrahlung and $\Gamma_{\rm NR}$ the one from the SM non-resonant contributions. The separation between the various processes is explained in Section \ref{sec:SM}. The four last lines correspond to possible 
new physics. Line $4$ is the contribution of the rescaled $Z$ peak, essentially as reported by the ATLAS and CMS collaborations, see Section \ref{sec:resc}; $\Gamma_{EFT}$ the contribution from the effective operator, Section \ref{sec:effOP}; $\Gamma_{UV62}$ the UV model  with masses $m_F =m_S =  62.5$ GeV; $\Gamma_{UV100}$  the same with $m_F =m_S =  100$ GeV.}
\label{tab:binsCMS}
\end{table}

\begin{table}[ht!]
\centering
\begin{tabular}{|c|c|c|c|c|c|c|c|c|}
\hline
  & \, 40-50 \,  & \, 50-60 \,   & \, 60-70 \,    & \, 70-80 \,   & \, 80-90 \,   & \, 90-100 \,   & \, 100-110 \,  & \, 110-125 \, \\
\hline
 $\Gamma_{\rm SM}$  & 0.013 & 0.013 & 0.013   & 0.021 & 0.131 & 0.325 & 0.024 & 0.240  \\
 \hline
 $\Gamma_{\rm tree}$   & 0.000 & 0.002 & 0.003   & 0.005 & 0.007 & 0.011 & 0.019 & 0.240  \\
\hline
$\Gamma_{\rm NR}$   & 0.012 & 0.008 & 0.005   & 0.003 & 0.002 & 0.001 & 0.000 & 0.000 \\
\hline
\hline
$\Gamma_{\rm resc}$  & 0.002 & 0.005 & 0.011   & 0.027 & 0.257 & 0.662 & 0.009 & 0.000  \\
\hline
$\Gamma_{\rm EFT}$  & 0.079 & 0.115 & 0.140   & 0.146 & 0.126 & 0.085 & 0.040 & 0.009  \\
\hline
$\Gamma_{\rm UV62}$   & 0.224 & 0.236 & 0.216   & 0.169 & 0.111 & 0.058 & 0.021 & 0.004 \\
\hline
$\Gamma_{\rm UV100}$  & 0.091 & 0.128 & 0.150   & 0.149 & 0.123 & 0.080 & 0.035 & 0.007  \\
\hline
\end{tabular}
\caption{For reference, the same as in Table \ref{tab:binsCMS}, but with no cuts.}
\label{tab:bins}
\end{table}

\subsection{A UV model with dark matter}

\label{sec:UV}

To explain the current data, the scale $\Lambda_R$ of the effective operator \eqref{eq:effOP} must be close to the electroweak scale (see also \cite{Boto:2023bpg}). This implies the presence of relatively light new degrees of freedom, raising concerns about the validity of an effective operator approach. Moreover, we have assumed that the dimension-8 operator \eqref{eq:effOP} is the sole indication of new physics and, accessorily, that $\Lambda_R$ is the same for the effective couplings to muons and electrons.
Altogether, it is possible that the new physics underlying \eqref{eq:effOP} has additional experimental implications. To at least partially address these concerns, we now consider a specific ultraviolet (UV) model, which is also related to the dark matter (DM) problem.

\begin{figure}[t]
	\begin{center}
		\subfigure[t][]{\includegraphics[width=0.23\textwidth]{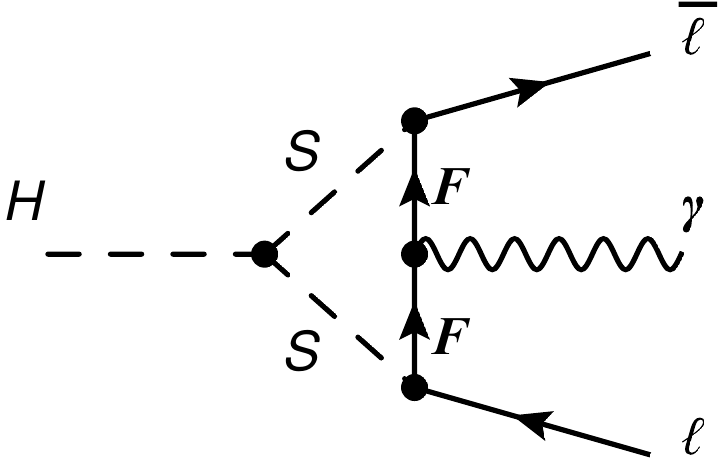}}
		\hspace{.6cm}
  	\subfigure[t][]{\includegraphics[width=0.23\textwidth]{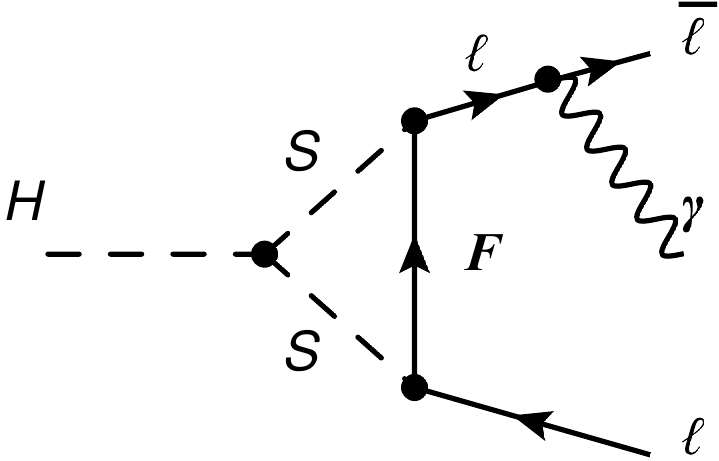}}
   		\hspace{.6cm}
   		\subfigure[t][]{\includegraphics[width=0.23\textwidth]{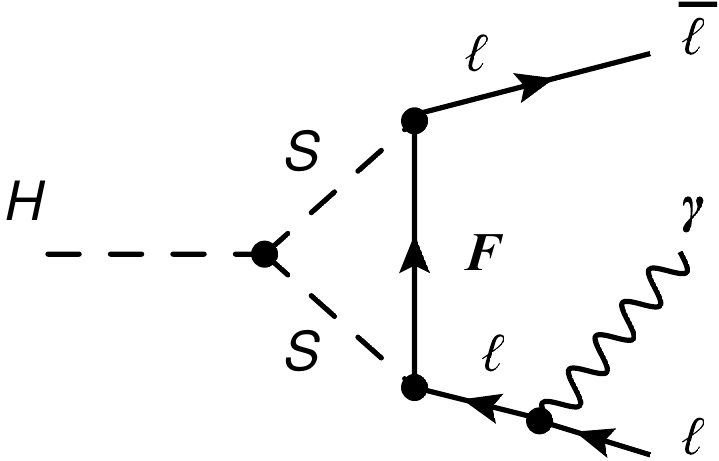}}
         \end{center}
	\caption{Amplitudes for the UV  model. Bremsstrahlung from final state leptons, see subfigures (b) and (c), is both loop and chirality suppressed. This contribution is negligible for both electrons and muons. In the chiral limit, amplitude (a), with the emission of the photon from a virtual fermion has the same structure as the effective operator \eqref{eq:effOP}. }
	\label{fig:LoopUV}
\end{figure}

An effective operator similar to \eqref{eq:effOP} has been proposed in \cite{Barger:2009xe} in the context of indirect searches for DM. In brief, the relevant process involves the annihilation of a spin-zero, \(S\)-wave, zero-velocity initial state (\(v \rightarrow 0\)) into dileptons and a gamma ray, \( S \rightarrow \ell^+\ell^- \gamma \). The initial state may correspond, for instance, to a pair of Majorana or scalar DM particles ($S$). For kinematic reasons, the annihilation of such DM particles into light fermions is helicity suppressed, with \( \sigma v \propto m_\ell^2 \). This suppression can be lifted by the emission of a gamma ray if the DM annihilation proceeds through a charged particle in the \( t \)-channel, a process known as virtual internal Bremsstrahlung, as illustrated in diagram (a) of Figure \ref{fig:LoopUV}. Furthermore, for kinematic reasons, the emitted gamma ray tends to be very energetic, with \( E_\gamma \sim m_{S} \), provided the DM particle mass and the mass of the particle in the \( t \)-channel are almost degenerate. Such features are particularly relevant for DM indirect searches \cite{Bergstrom:1988fp, Barger:2009xe, Bringmann:2012vr, Toma:2013bka,Giacchino:2013bta,Giacchino:2014moa}.

Motivated by these considerations, we consider the following  model \cite{Toma:2013bka,Giacchino:2013bta}
\begin{equation}
    \mathcal{L} \supset {1\over2} \partial_\mu S \partial^\mu S - {1\over 2} m_S^2 S^2 + \bar{F} (i \slashed{D} - m_{F})F - \sum_\ell (y_\ell S \bar{F} \ell_{R} + h.c.) - \frac{\lambda_{hs}}{2} S^2 |\Phi|^2 
    \label{eq:LagUV}
\end{equation}
where  $S$ is a real scalar, singlet of the SM. Potentially, $S$ is a DM candidate \cite{Silveira:1985rk,McDonald:1993ex,Burgess:2000yq}. It couples to the SM  through the Higgs portal \cite{Patt:2006fw} and also, through Yukawa interactions, to the (right-handed) leptons via a leptonic vector-like fermion $F$, with hypercharge $Y  = Q =-1$.  For the purpose of DM phenomenology, the Lagrangian is built to be invariant under a $Z_2$ symmetry with $S$ and $F$ odd and the SM field even. There is therefore no mixing between $F$ and SM leptons and no source of lepton flavour violation. For DM stability, we  assume that the $Z_2$ symmetry is not spontaneously broken, $\langle S\rangle \neq 0$. Hence, there is no mixing between the Higgs and the $S$ particle. 

Given this Lagrangian, the new particles may directly contribute to the decay \( H \rightarrow \ell^+ \ell^- \gamma \) through the diagrams shown in Figure \ref{fig:LoopUV}. Importantly, at one loop, the model does not lead to deviations of either $H \rightarrow \gamma^{(\ast)} \gamma$ or $H \rightarrow Z \gamma$.
 However, both to explain the observed excess of events and to avoid significant lepton flavour violation, we must assume that the Yukawa couplings \( y_\ell \) are the same for electrons and muons. Clearly, this is  not natural so the Lagrangian \eqref{eq:LagUV} is a toy model, which we propose for the sake of illustration. That being said, its phenomenology depends only on 4 free parameters, the two masses $m_S$ and $m_F$, the Yukawa coupling $y_\ell = y_e = y_\mu$ and the quartic coupling $\lambda_{hs}$. We first focus on the process $H \rightarrow \ell^+\ell^- \gamma$. The concrete implications, as well as other experimental and observational constraints, are briefly discussed in the next section.  

The general expression for the amplitude of Figure \ref{fig:LoopUV}  is not particularly transparent. However, it takes a relatively simpler form in the  chiral limit $m_\ell \rightarrow 0$, which is an excellent approximation given the smallness of the electron and muon masses, $m_\ell \ll m_H$. In that limit, the Bremsstrahlung from final states dileptons is identically vanishing, a result which is similar to the behavior of the DM annihilation amplitude in the very same model, see  \cite{Barger:2009xe,Giacchino:2014moa} for details. Thus the remaining amplitude stems only from the box diagram (a) in Figure \ref{fig:LoopUV} and has, as expected,  the structure of eq. \eqref{eq:SMloop} with form factors  $\Delta B_{1,2}$, which are function of $m_S$, $m_F$ and $m_H$. Their explicit expression in terms of Passarino-Veltman functions is given in the Appendix \ref{app:FF}. For $M = m_S, m_F \gg m_H$, the amplitude reduces to that of the effective operator, with $1/\Lambda_R^4 \propto y_{\ell}^2 \lambda_{hs}/M^4$. However, to explain the excess of events observed at the LHC, the mass scales must be close to the electroweak scale. So, the mass of the $H$ cannot be neglected and we use the full analytical expressions given in  Appendix \ref{app:FF}.

\subsection{$H$ decay signature and constraints}

\label{sec:constraints}

\begin{figure}[]
{\includegraphics[width=0.99\textwidth]{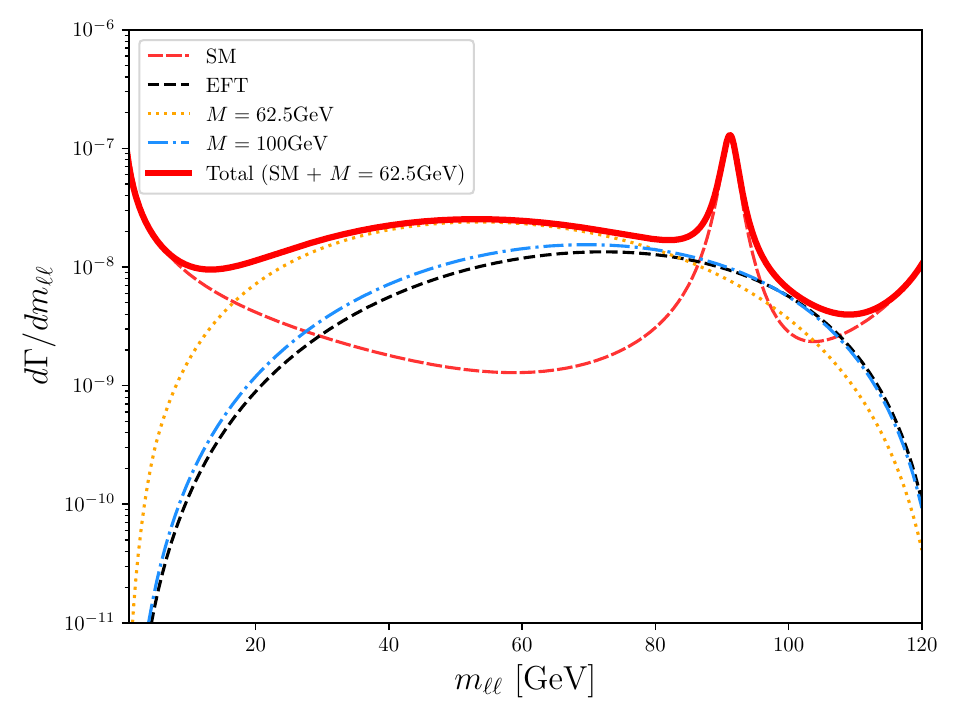}}
\caption{Possible UV complete contributions to the differential decay rate for two benchmark candidates, $M = m_S = m_F = 62.5$ GeV (dotted orange) and $M = 100$ GeV (dot-dashed blue). The black dashed curve corresponds to the effective operator, as in Figure \ref{fig:rateEffOp}. Total contribution (red solid line) corresponds to the sum of the SM and $M = 62.5$ GeV model. In this figure, no cuts are assumed. }
    \label{fig:rate_NP2}
\end{figure}

\subsubsection{Excess in $H \rightarrow \ell^+ \ell^- \gamma$}
\label{sec:excessUV}

From the explicit expressions provided by  the UV model, we can verify that the  differential decay rate is indeed similar to the prediction of the effective operator \eqref{eq:effOP}. However, its shape is more or less skewed toward smaller values depending on the choice of $m_S$ and $m_F$ (see also Appendix \ref{app:DM_decay}). 
We provide two examples of this behaviour in Figure \ref{fig:rate_NP2}. In both cases, the scalar and fermion masses are chosen to be almost degenerate, the scalar $S$ being assumed to be just slightly  lighter,  \( m_S \lesssim m_F = 62.5 \) GeV, just above $m_H/2$  (dot-dashed green line) and  \( m_S \lesssim m_F = 100 \) GeV (dot-dashed blue), together with the shape for the effective operator, which is the same as in Figure \ref{fig:rateEffOp}. All the curves have been normalized in such a way as to reproduce the excess, given the experimental cuts. We assumed that $m_S > m_H/2$  to avoid the constraint from invisible \( H \) decay \cite{ParticleDataGroup:2024cfk}. 

    Similar to the discussion of the previous section on the effective operator, we show in the last column of Table \ref{tab:sample} the signal strength for the benchmark model $M = m_S \gtrsim m_F = 62.5$ GeV, which depicts  the largest deformation compared to the spectrum from  the effective operator. Accordingly, the values of the signal strength for the other benchmark model $M = m_S \gtrsim m_F = 100$ GeV are similar to that of the effective operator. The binned decay rates are clearly more sensitive probes of possible deviations, both from the case of a rescaled $Z$ peak and of the effective operator. The values are given in the last two lines of Tables \ref{tab:binsCMS} and  \ref{tab:bins} for both benchmark models.

The other reasons for choosing these two benchmarks and, in particular, for assuming a compressed mass spectrum $m_S \lesssim m_F$ are as follows. First, the new particles must be as light as possible to give  a significant excess. After all, the models leads to a dimension 8 operator, which scales like $1/\Lambda_R^4 \sim  \lambda_{hs} y_{\ell}^2/M^4$. The  combination of couplings required is  $\lambda_{hs} y_{\ell}^2 = 0.72$  for $m_S = m_F = 62.5$ GeV. For instance,  $y_{\ell} = 1.66$ for $\lambda_{hs}= \lambda = 0.26$ with $\lambda$ the SM Higgs quartic coupling. For $m_S = m_F = 100$ GeV, however, the product of couplings is already large, $\lambda_{hs} y_{\ell}^2 = 28.1$. Assuming $y_{\ell}^2 \lesssim 4 \pi$, this corresponds to $\lambda_{hs} \gtrsim 2.2$. Clearly,  larger mass values  are essentially excluded and the particles are constrained to lie between $m_H/2$ and  about $100 \mbox{\rm GeV}$. Our last motivation for assuming a compressed spectrum is to  avoid current collider and DM direct detection constraints. These are considered below, together with constraints from electroweak precision tests. Speaking of DM, there are other motivations to consider a compressed mass spectrum, related to the possible indirect signatures of a decaying DM particle. To avoid a lengthy digression, this aspect is relegated to Appendix \ref{app:DM_decay}. 

% \begin{figure}[h]
% \centering
% \begin{tikzpicture}
% \begin{feynman}
%   \vertex (i) at (-2,0) {\(\mu\)};
%   \vertex (f) at (2,0) {\(\mu\)};
%   \vertex (v1) at (-1,0);
%   \vertex (v2) at (1,0);
%   \vertex (v3) at (0,1);
%   \vertex (p) at (0,-1);
%   \vertex (q) at (0,0);

%   \diagram* {
% %   
%  (i) -- [plain,thick] (v1) -- [scalar, edge label =\(S\), half left] (v2) ,
%     (v2) -- [plain,thick] (f),
%     (v1) -- [plain, thick,edge label =\(F\)] (q) -- [plain, thick, edge label =\(F\)] (v2) ,
%    };

%   % Connect the photon to the fermion loop
%   \draw[photon] (p) -- (0,0);
% \end{feynman}
% \end{tikzpicture}
% \caption{UV model contribution to the muon anomalous  magnetic moment.}
% \label{fig:muB}
% \end{figure}

\begin{figure}[t]
	\begin{center}
		\subfigure[t][]{\includegraphics[width=0.23\textwidth]{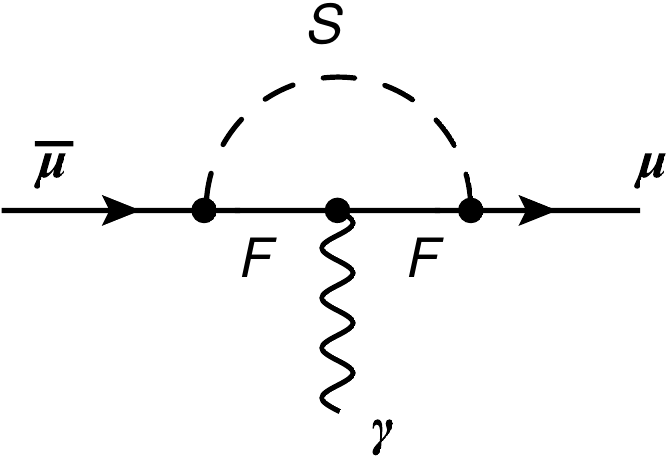}}
         \end{center}
	\caption{UV model contribution to the muon anomalous  magnetic moment.}
	\label{fig:muB}
\end{figure}

\subsubsection{Muon magnetic dipole moment}

The UV model can also affect the  muon anomalous magnetic dipole moment $g$, see figure \ref{fig:muB}. From this amplitude, we get that its contribution to the muon electromagnetic form factor $F_2(q^2)$ with $q$ the photon moment, see e.g. \cite{Peskin:1995ev}, is such that 
\begin{equation}
    \Delta F_2(0)\, \hat = \,\Delta\left({g_\mu-2\over 2}\right)\,\hat =\, \Delta a_\mu =  {y_l^2 \over 192 \pi^2} {m_\mu^2\over M^2} 
    \label{eq:muonMM}
\end{equation}
is positive. 
Here, $m_\mu$ is the muon mass and we have set $M = m_S = m_F$ to compel with our calculations of the $H \rightarrow l^+ l^- \gamma$ process. For instance, for $y_l = 1.28$ and $m_S = 62.5$ GeV (corresponding to $\lambda_{hs} y_l^2 = 0.72$  if $\lambda_{hs}= 0.44$, see section \ref{sec:excessUV}), \eqref{eq:muonMM} gives $\Delta a_\mu = 2.5 \cdot 10^{-9}$. Similarly, we can get the same shift with $y_l = 2.05$ for $m_S = 100$ GeV, corresponding respectively to $\lambda_{hs} y_l^2 = 28.1$ for $\lambda_{hs} = 6.7$. The  status of the theoretical prediction for the muon anomalous magnetic moment within the SM is currently unclear \cite{Szabo:2024pjz}. Yet, it remains of interest that the deviation of $a_\mu$ reported by the Muon $g-2$ collaboration \cite{Muong-2:2021ojo} is  $a_\mu({\rm exp}) - a_\mu({\rm SM}) = (251 \pm 59) \cdot 10^{-11}$, within the range of the UV model.

\subsubsection{Electroweak precision tests (EWPT)}

The vectorlike lepton $F \sim (1,1,-1)$ is rather light and couples  both to the photon and the $Z$ boson. It may thus contribute to oblique parameters \cite{Peskin:1990zt}. As the benchmark candidates we consider are comparable to $m_Z$ ($m_F \lesssim m_Z$), we cannot rely on  the standard $S,T,U$ parameters and must use instead the extended set of oblique parameters ($S, \ldots, X$) of \cite{Maksymyk:1993zm}. From the expressions given in \cite{Albergaria:2023nby}, where the authors study the oblique parameters for general vector-like EW representations (see also \cite{Moreno:1991mf}), we have
\begin{equation}
     S = - U = -0.013 \,(-0.0042), \quad  V = - 0.026\, (-0.0065
    )\quad \mbox{\rm and}\quad  X = 0.014 \,(0.0045)
\end{equation} for our benchmarks $m_F = 62.5 \,(100)$ GeV.
The contributions of $F$ to $T$ (custodial symmetry breaking) and $W$ (contribution to $W$ decay) are identically vanishing. The values quoted are quite small compared to typical uncertainties on the oblique parameters, which are ${\cal O}(0.1)$ \cite{ParticleDataGroup:2024cfk}. To fully assess the effect of $F$, we should perform a global fit to EW precision measurements. Here, instead, we just quote three contributions that involve the $S,U, V$ and $X$ oblique parameters:  $Z$ invisible ($\Gamma_{\rm inv}$) and charged lepton ($\Gamma_{\rm \ell\ell}$) decay,  and to the $W$ mass (see Table I in \cite{Maksymyk:1993zm}),
\begin{eqnarray}
 \left.{\Delta \Gamma\over \Gamma}\right\vert_{\rm inv} &=& 0.0078 \,\vert V\vert = 0.0002 \,(0.00005) \, < 0.003
%\\
\\ \left.{\Delta \Gamma\over \Gamma}\right\vert_{\rm \ell\ell} &=& \vert -0.0021\, S - 0.0044\, X + 0.0078\, V\vert  = 0.0002 \,(0.00006) \, < 0.001\\
\Delta m_W &=&  ( -0.29 \,S  + 0.34\, U)\, \mbox{\rm GeV} = 0.0082 \,(0.0026)  \, \mbox{\rm GeV}\,  < 0.013 \, \mbox{\rm GeV} 
\end{eqnarray}
which show that the oblique corrections are significantly smaller than the current $1\sigma$ experimental uncertainties \cite{ParticleDataGroup:2024cfk}.

\subsubsection{Constraints from colliders}

We now consider pair production of $F \bar F$ at the LHC. The $F$ and $S$ are assumed to be nearly degenerate, with $m_F \gtrsim m_S$. We assume that the mass difference is large enough to allow for decay into electrons and muons, but not much more. As the Yukawa coupling is not tiny, the process $\overset{\scriptscriptstyle(-)}{F}  \rightarrow S + \overset{\scriptscriptstyle(-)}{\ell} $ leads to  soft leptons, which escape detection. The production of $F \bar F$ is thus equivalent to missing energy, $pp \rightarrow $ jets $ + \slash \!\!\!\!E$. 

It could be of interest to analyse  constraints on such scenario more in depth.
Here, to put constraints on $m_F$, we have simply recasted the limits from missing energy searches provided by the CMS Collaboration. 
Concretely, considering a $Z'$ mediator with a universal coupling $g'=0.25$ to quarks and a coupling $g=1$ to  a fermionic DM candidate $\chi$, a bound $m_\chi > 94$ GeV has been set for the case of $m_{Z'} = m_Z$ \cite{cmscollaboration2024darksectorsearchescms}. In our setup, the $F \bar F$ can be produced through the SM $Z$ or a $\gamma^\ast$ and this, with known couplings. To recast the limit, we have  compared the cross sections of the two processes $q \overline{q} \rightarrow Z' \rightarrow \chi \overline{\chi}$ and $q \overline{q} \rightarrow Z/\gamma \rightarrow F \overline{F}$ and have determined the value of the outgoing fermion mass $m_F$ for the $\gamma/Z$ process that corresponds to the same cross section as the one for the $Z'$ process with a mass $m_\chi$ of 94 GeV for the outgoing fermion. For fixed energy, the cross section through the $Z'$ is larger than the one through the $Z$ and $\gamma^\ast$, so the bound on $F$ should lie at a lower mass $m_F$.  We did our recasting both with and without the phase space suppression factor for the production of the $F \bar F$. The bound on $m_F$ we found agrees to within 5\%, $m_F \gtrsim 67$ GeV. The  benchmark candidate with $m_F \gtrsim m_S = 62.5$ GeV is perhaps dangerously close to our estimated bound. The one with $m_F \gtrsim m_S = 100$ GeV is  on the safe side.  

\subsubsection{Comments on DM direct detection}

The last phenomenological aspect we briefly consider takes seriously the possibility that the scalar particle $S$ could be a DM candidate, even if only as a subdominant component.

There is a vast literature on a singlet scalar as a possible DM candidate, starting with \cite{Silveira:1985rk,McDonald:1993ex,Burgess:2000yq}. In such scenario, the $S$ interacts with the SM sector solely through the Higgs portal, with coupling $\lambda_{hs}$. For thermal freeze-out of massive particles, the abundance (here $Y_S = n_S/s$ with $s$ the entropy density) is inversely proportional to the particle annihilation cross section  $Y \propto 1/\langle \sigma v\rangle \propto 1/\lambda_{hs}^2$. The locus of DM candidates can thus be reported in the plane $\lambda_{hs}-m_S$, as in the Figure 2 of \cite{Arcadi:2024wwg}. In the same plane, one can report constraints from direct detection searches, as the DM-nucleon collision cross section is $\sigma_{S-n} \propto \lambda_{hs}^2$. The same plot can be used to set constraints on a subdominant form of dark matter. Indeed, as the decay rate for DM to nucleon collisions, $\propto \sigma_{S-n} n_S$, is essentially independent of $\lambda_{hs}$, the plot remains the same provided the locus is taken to correspond to a fraction $f_S = n_S/n_{\rm dm}$ of $S$ particles compared to the observed DM and the vertical axis is read to correspond to $\lambda_{hs} f_S^{1/2}$ instead of $\lambda_{hs}$. 

For instance, for the benchmark $S$ particle, with mass $m_S = 100$ GeV, the quartic coupling must be such that $\lambda_{hs} \gtrsim 2.2$. This is  to be compared to the required coupling $\lambda_{hs} \approx 0.04$ for $f_S = 1$. If stable, the $S$ particle is thus a subdominant form of DM, with an abundance that is  suppressed by a factor $f_S \approx (0.04/2.2)^2 \approx 3\cdot 10^{-4}$. Yet, it is excluded by DM direct detection constraints, at least {\em a priori}. The ligther benchmark particle $S$ with mass only slightly above $m_H/2$, could still escape the direct detection constraints, being close to the $H$ resonance, see again Figure 2 of \cite{Arcadi:2024wwg}. All these considerations however neglect the fact that the $S$ particle also interacts with the SM through Yukawa interactions with SM leptons through the $F$ particle  \cite{Toma:2013bka,Giacchino:2014moa}, an instance of DM through a t-channel interaction (see e.g. \cite{Arina:2023msd}). The implications for dark matter phenomenology are therefore much richer; however, a detailed discussion is probably unnecessary and, in any case, beyond our scope, as our aim was to be illustrative.

\section{Conclusions}

In this work, motivated by indications of an excess reported by both the ATLAS and CMS collaborations, we have explored the possibility that a new non-resonant process could contribute to the background for measurements of the effective Higgs coupling to $Z\gamma$. We have formulated this new physics scenario both in terms of an effective operator (eq.~\eqref{eq:effOP}) and, given that the induced new physics scale is close to the electroweak scale, using a simplified UV model (eq.~\eqref{eq:LagUV}), which is also motivated by the dark matter problem.  Our key results are illustrated in Figures~\ref{fig:rate_NP1}, \ref{fig:rateEffOp}, and \ref{fig:rate_NP2}, which depict the differential decay rate $d\Gamma(H \rightarrow l^+l^- \gamma)/dm_{\ell\ell}$ as a function of the dilepton invariant mass. Given the current cuts applied by the experiments, the non-resonant new physics contributions lead to broad, bumpy features, as shown, for instance, in Figure~\ref{fig:rateEffOp}. These contributions are normalized so as to produce the same excess of events as the rescaled $Z$-peak contribution shown in Figure~\ref{fig:rate_NP1}.

While an excess from the $Z$ peak is largely independent of the applied $m_{\ell\ell}$ cuts, using more stringent cuts leads to a decrease in the non-resonant contribution; see Table~\ref{tab:sample}. Such cuts could thus result in the disappearance of the excess. This, in turn, could be mistakenly interpreted as the disappearance of a statistical fluctuation in previous experimental results, whereas, in reality, it may instead be suppressing a genuine hint of off-$Z$ new physics.  Distinguishing between these two possibilities is, in principle, a matter of measuring the differential cross section in bins of $m_{\ell\ell}$, particularly around and below the $Z$ boson mass. We encourage the ATLAS and CMS collaborations to implement this refinement in their future searches in the $\ell^+ \ell^- \gamma$ final state. 

The UV model we have considered for illustrative purposes is consistent with experiment but is, in its present form, rather artificial. It is fine-tuned, and we have had to assume several unrealistic features—the most important being a rather low new physics scale, a compressed mass spectrum $m_S \approx m_F$, and identical Yukawa couplings for electrons and muons. 
Even though we find it unlikely that such new physics  can {\em significantly} contribute to the process $H \rightarrow \ell^+\ell^- \gamma$, for completeness, we have examined additional experimental constraints that could be imposed on such a model, including the possibility that one of the new particles we invoked could constitute a subdominant form of dark matter; see Section~\ref{sec:constraints}. Looking ahead, our statement and even the model itself should remain relevant for future analyses, even if the observed excess were to disappear.

\section*{Acknowledgments}
We thank Giorgio Arcadi, Debtosh Chowdhury and Laura Lopez Honorez for discussions. We also thank Ivan Nišandžić for providing the code we used to cross-check parts of our work. This work has been supported by the FRS/FNRS, the FRIA, {the BLU-ULB Brussels laboratory of the Universe} and by the IISN convention No. 4.4503.15.

\begin{appendix}
\section{Form factors}
\label{app:FF}

In the chiral limit, $m_\ell \rightarrow 0$, the form factors corresponding to the UV model of Section \ref{sec:UV}, expressed in terms of Pasarino-Veltman functions (see \cite{Kachanovich:2020dah} for conventions),  are given by 
\begin{eqnarray}
    \nonumber \Delta B_{1} &=& \frac{3 \pi^2 e^2 Y^2 m_{H}^{2}}{4 m_{W} \sin \theta_W} \\
     \nonumber &\times&   \Bigg\{  \frac{3}{t\, (s + t)}\Big[ B_{0}(t,  m_{F}^{2},  m_{S}^{2}) -  B_{0}(m_{H}^{2},  m_{S}^{2},  m_{S}^{2}) \Big]  \\
    \nonumber && - \frac{2(m_{F}^{2} - m_{S}^2)}{s\, t^2 (s+t)} \Big[ (s^2+2s \,t + t^2) C_{0}(0, m_{H}^{2}, u,  m_{F}^{2},  m_{S}^{2},  m_{S}^{2}) \\
    \nonumber && \quad + (s + t) \big( u\, C_{0}(0, 0, u,  m_{F}^{2},   m_{F}^{2},  m_{S}^{2}) - (s + u) C_{0}(0, t, m_{H}^{2},    m_{S}^{2}, m_{F}^{2},  m_{S}^{2}) \big)
    \\
    \nonumber && \quad + t\, (s + t) C_{0}(0, 0, t, m_{F}^{2}, m_{F}^{2},  m_{S}^{2}) \Big] \\
     \nonumber && - \frac{2}{s\, t^2} \Big[ 2 m_{F}^{2} m_{S}^{2} m_{H}^{2} - m_{F}^{4} m_{H}^{2} - m_{F}^{2} m_{H}^{2} t - m_{S}^{4} m_{H}^{2} + m_{S}^{2} m_{H}^{2} t \\
     \nonumber && \quad + m_{F}^{4} s - 2 m_{F}^{2} m_{S}^{2} s - m_{F}^{2} s\,t + m_{F}^{2} t^2 + m_{S}^{4} s - m_{S}^{2} s \, t - m_{S}^{2} t^2 \Big] \\
     && \times D_{0} (0, 0, 0, m_{H}^{2},t, u, m_{S}^{2}, m_{F}^{2}, m_{F}^{2}, m_{S}^{2}) \Bigg\}
\end{eqnarray}
while the form factor $\Delta B_{2}$ is obtained from the above expression by replacing $t = (p_1+q)^2$ by $u = (p_2 + q)^2$, with $p_1$ ($p_2$) the four-momentum of the lepton (resp. antilepton) and $q$ that of the photon (see also eq.~\eqref{eq:SMloop}. In the chiral approximation, the amplitude stems entirely from the box diagram (a) of Figure \ref{fig:LoopUV}.

{For numerical evaluations, we used the \emph{Collier} library \cite{Denner:2016kdg, Denner:2005nn, Denner:2010tr} via the Wolfram \emph{Mathematica} \cite{Mathematica} interface \emph{CollierLink} \cite{Patel:2015tea, Patel:2016fam} as well as \emph{Python}. Additional numerical checks were performed using the \emph{LoopTools} software package \cite{Hahn:1998yk}. Dirac algebra manipulations and Passarino-Veltman reductions to scalar loop functions were carried out with \emph{FeynCalc} \cite{Shtabovenko:2023idz, Shtabovenko:2016sxi, Shtabovenko:2020gxv, Mertig:1990an}. For certain intermediate checks, we utilized \emph{Package-X} \cite{Patel:2015tea}, which was linked to \emph{FeynCalc} via the \emph{FeynHelpers} \cite{Shtabovenko:2016whf} interface.}

\section{Comments on decaying scalar DM}
\label{app:DM_decay}

\begin{figure}[]
{\includegraphics[width=0.9\textwidth]{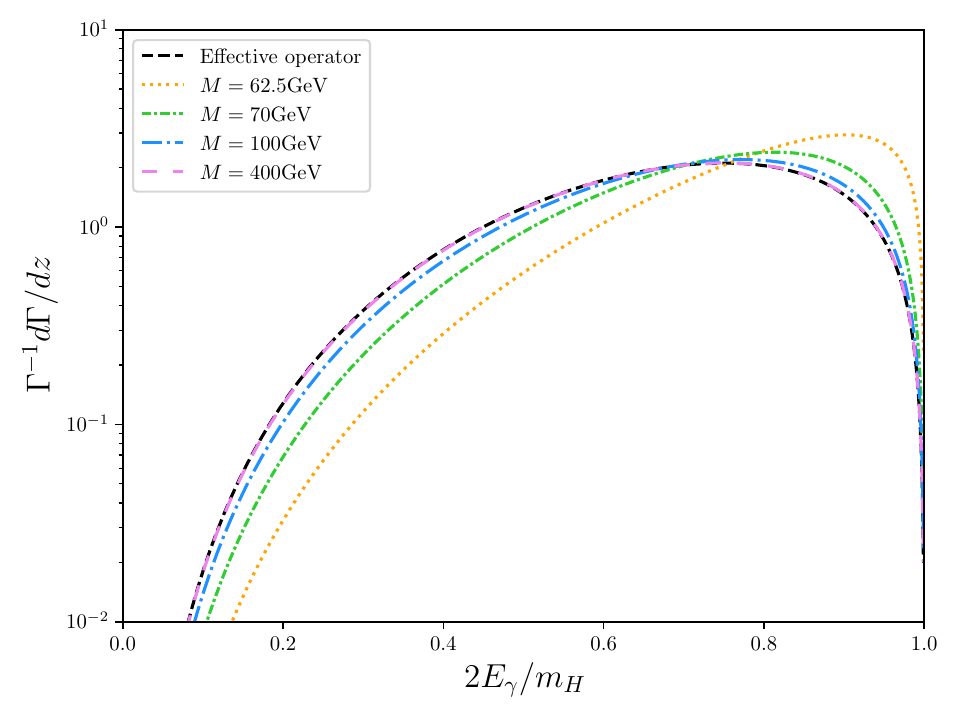}}
	\caption{Gamma-ray spectral signature for various choices of $m_S = m_F = M$. The peak that is typical of VIB (see text) in the case of DM annihilation is particularly visible for $m_S \approx m_F \approx m_X/2$, but is smeared due to integration over the loop momentum. For large $m_S,m_F \gg m_X$, the spectrum becomes indistinguishable from that of the effective operator.}
	\label{fig:Xdecay}
\end{figure}

The Lagrangian \eqref{eq:LagUV} given in Section \ref{sec:UV} describes a dark matter (DM) model that was introduced in \cite{Toma:2013bka,Giacchino:2013bta}. The DM candidate is a real scalar $S$, whose annihilation cross-section into dileptons is d-wave and chirally suppressed, $\langle \sigma v \rangle \propto (v^4,m_l^2)$. This suppression is lifted by radiative corrections, in particular through the emission of a $\gamma$ ray by the charged fermion $F$ in the t-channel—a process known as virtual internal Bremsstrahlung (VIB), see also \cite{Bergstrom:1988fp} for the case of Majorana DM. Furthermore, the emitted gamma-ray tends to peak at the DM mass if $m_S \sim m_F$, a spectral feature of interest for indirect DM searches \cite{Bringmann:2012vr}. 

This model is also relevant for the case of a decaying DM particle \cite{Barger:2009xe}, with  $S$ and $F$ in loop diagrams, as in Figure \ref{fig:LoopUV}, with some scalar DM particle, say $X$, replacing the role of the Higgs. In \cite{Barger:2009xe}, the decay of $X \rightarrow \ell^+\ell^- \gamma$ was treated at the level of the an effective operator, akin to Eq.~\eqref{eq:effOP}, arguing that the spectral signature of the $\gamma$-ray should be independent of the properties of the particle in the loop. While this is clearly the case if $m_{F,S} \gg m_X$, our calculations show that the gamma-ray spectrum shape is distinct from that of the effective operators if $m_{S,F,X}$ are of the same order. We depict some instances of such behavior in Figure \ref{fig:Xdecay} which gives the  partial decay rate  (normalized to the total decay rate) in function of $z=E_\gamma/m_X$ for some choices of $m_F=m_S$. The benchmark values are given for $m_X = m_H$. Compared to the case of DM annihilation, the spectra are smeared, as result of the integration on the loop momenta, but tends to peak towards $E_\gamma \approx m_X/2$. Such features are just complementary to the ones observed in the dilepton invariant mass spectrum, as discussed in the bulk of our work, compare e.g. the dashed green curve in Figure \ref{fig:Xdecay} with the one in Figure \ref{fig:rate_NP1}.

% \begin{figure}[]
% {\includegraphics[width=0.99\textwidth]{Plots/ml62mf10to600.pdf}}
% 	\caption{}
% 	\label{fig:Gamma_1}
% \end{figure}

\end{appendix}
\bibliographystyle{apsrev4-1}

\bibliography{main}

%merlin.mbs apsrev4-1.bst 2010-07-25 4.21a (PWD, AO, DPC) hacked
%Control: key (0)
%Control: author (72) initials jnrlst
%Control: editor formatted (1) identically to author
%Control: production of article title (-1) disabled
%Control: page (0) single
%Control: year (1) truncated
%Control: production of eprint (0) enabled
\begin{thebibliography}{72}%
\makeatletter
\providecommand \@ifxundefined [1]{%
 \@ifx{#1\undefined}
}%
\providecommand \@ifnum [1]{%
 \ifnum #1\expandafter \@firstoftwo
 \else \expandafter \@secondoftwo
 \fi
}%
\providecommand \@ifx [1]{%
 \ifx #1\expandafter \@firstoftwo
 \else \expandafter \@secondoftwo
 \fi
}%
\providecommand \natexlab [1]{#1}%
\providecommand \enquote  [1]{``#1''}%
\providecommand \bibnamefont  [1]{#1}%
\providecommand \bibfnamefont [1]{#1}%
\providecommand \citenamefont [1]{#1}%
\providecommand \href@noop [0]{\@secondoftwo}%
\providecommand \href [0]{\begingroup \@sanitize@url \@href}%
\providecommand \@href[1]{\@@startlink{#1}\@@href}%
\providecommand \@@href[1]{\endgroup#1\@@endlink}%
\providecommand \@sanitize@url [0]{\catcode `\\12\catcode `\$12\catcode `\&12\catcode `\#12\catcode `\^12\catcode `\_12\catcode `\%12\relax}%
\providecommand \@@startlink[1]{}%
\providecommand \@@endlink[0]{}%
\providecommand \url  [0]{\begingroup\@sanitize@url \@url }%
\providecommand \@url [1]{\endgroup\@href {#1}{\urlprefix }}%
\providecommand \urlprefix  [0]{URL }%
\providecommand \Eprint [0]{\href }%
\providecommand \doibase [0]{http://dx.doi.org/}%
\providecommand \selectlanguage [0]{\@gobble}%
\providecommand \bibinfo  [0]{\@secondoftwo}%
\providecommand \bibfield  [0]{\@secondoftwo}%
\providecommand \translation [1]{[#1]}%
\providecommand \BibitemOpen [0]{}%
\providecommand \bibitemStop [0]{}%
\providecommand \bibitemNoStop [0]{.\EOS\space}%
\providecommand \EOS [0]{\spacefactor3000\relax}%
\providecommand \BibitemShut  [1]{\csname bibitem#1\endcsname}%
\let\auto@bib@innerbib\@empty
%</preamble>
\bibitem [{\citenamefont {Cahn}\ \emph {et~al.}(1979)\citenamefont {Cahn}, \citenamefont {Chanowitz},\ and\ \citenamefont {Fleishon}}]{Cahn:1978nz}%
  \BibitemOpen
  \bibfield  {author} {\bibinfo {author} {\bibfnamefont {R.~N.}\ \bibnamefont {Cahn}}, \bibinfo {author} {\bibfnamefont {M.~S.}\ \bibnamefont {Chanowitz}}, \ and\ \bibinfo {author} {\bibfnamefont {N.}~\bibnamefont {Fleishon}},\ }\href {\doibase 10.1016/0370-2693(79)90438-6} {\bibfield  {journal} {\bibinfo  {journal} {Phys. Lett. B}\ }\textbf {\bibinfo {volume} {82}},\ \bibinfo {pages} {113} (\bibinfo {year} {1979})}\BibitemShut {NoStop}%
\bibitem [{\citenamefont {Bergstrom}\ and\ \citenamefont {Hulth}(1985)}]{Bergstrom:1985hp}%
  \BibitemOpen
  \bibfield  {author} {\bibinfo {author} {\bibfnamefont {L.}~\bibnamefont {Bergstrom}}\ and\ \bibinfo {author} {\bibfnamefont {G.}~\bibnamefont {Hulth}},\ }\href {\doibase 10.1016/0550-3213(85)90302-5} {\bibfield  {journal} {\bibinfo  {journal} {Nucl. Phys. B}\ }\textbf {\bibinfo {volume} {259}},\ \bibinfo {pages} {137} (\bibinfo {year} {1985})},\ \bibinfo {note} {[Erratum: Nucl.Phys.B 276, 744--744 (1986)]}\BibitemShut {NoStop}%
\bibitem [{\citenamefont {Spira}\ \emph {et~al.}(1992)\citenamefont {Spira}, \citenamefont {Djouadi},\ and\ \citenamefont {Zerwas}}]{Spira:1991tj}%
  \BibitemOpen
  \bibfield  {author} {\bibinfo {author} {\bibfnamefont {M.}~\bibnamefont {Spira}}, \bibinfo {author} {\bibfnamefont {A.}~\bibnamefont {Djouadi}}, \ and\ \bibinfo {author} {\bibfnamefont {P.~M.}\ \bibnamefont {Zerwas}},\ }\href {\doibase 10.1016/0370-2693(92)90331-W} {\bibfield  {journal} {\bibinfo  {journal} {Phys. Lett. B}\ }\textbf {\bibinfo {volume} {276}},\ \bibinfo {pages} {350} (\bibinfo {year} {1992})}\BibitemShut {NoStop}%
\bibitem [{\citenamefont {Aad}\ \emph {et~al.}(2020)\citenamefont {Aad} \emph {et~al.}}]{ATLAS:2020qcv}%
  \BibitemOpen
  \bibfield  {author} {\bibinfo {author} {\bibfnamefont {G.}~\bibnamefont {Aad}} \emph {et~al.} (\bibinfo {collaboration} {ATLAS}),\ }\href {\doibase 10.1016/j.physletb.2020.135754} {\bibfield  {journal} {\bibinfo  {journal} {Phys. Lett. B}\ }\textbf {\bibinfo {volume} {809}},\ \bibinfo {pages} {135754} (\bibinfo {year} {2020})},\ \Eprint {http://arxiv.org/abs/2005.05382} {arXiv:2005.05382 [hep-ex]} \BibitemShut {NoStop}%
\bibitem [{\citenamefont {Tumasyan}\ \emph {et~al.}(2023)\citenamefont {Tumasyan} \emph {et~al.}}]{CMS:2022ahq}%
  \BibitemOpen
  \bibfield  {author} {\bibinfo {author} {\bibfnamefont {A.}~\bibnamefont {Tumasyan}} \emph {et~al.} (\bibinfo {collaboration} {CMS}),\ }\href {\doibase 10.1007/JHEP05(2023)233} {\bibfield  {journal} {\bibinfo  {journal} {JHEP}\ }\textbf {\bibinfo {volume} {05}},\ \bibinfo {pages} {233} (\bibinfo {year} {2023})},\ \Eprint {http://arxiv.org/abs/2204.12945} {arXiv:2204.12945 [hep-ex]} \BibitemShut {NoStop}%
\bibitem [{\citenamefont {Aad}\ \emph {et~al.}(2024)\citenamefont {Aad} \emph {et~al.}}]{ATLAS:2023yqk}%
  \BibitemOpen
  \bibfield  {author} {\bibinfo {author} {\bibfnamefont {G.}~\bibnamefont {Aad}} \emph {et~al.} (\bibinfo {collaboration} {ATLAS, CMS}),\ }\href {\doibase 10.1103/PhysRevLett.132.021803} {\bibfield  {journal} {\bibinfo  {journal} {Phys. Rev. Lett.}\ }\textbf {\bibinfo {volume} {132}},\ \bibinfo {pages} {021803} (\bibinfo {year} {2024})},\ \Eprint {http://arxiv.org/abs/2309.03501} {arXiv:2309.03501 [hep-ex]} \BibitemShut {NoStop}%
\bibitem [{\citenamefont {Barducci}\ \emph {et~al.}(2023)\citenamefont {Barducci}, \citenamefont {Di~Luzio}, \citenamefont {Nardecchia},\ and\ \citenamefont {Toni}}]{Barducci:2023zml}%
  \BibitemOpen
  \bibfield  {author} {\bibinfo {author} {\bibfnamefont {D.}~\bibnamefont {Barducci}}, \bibinfo {author} {\bibfnamefont {L.}~\bibnamefont {Di~Luzio}}, \bibinfo {author} {\bibfnamefont {M.}~\bibnamefont {Nardecchia}}, \ and\ \bibinfo {author} {\bibfnamefont {C.}~\bibnamefont {Toni}},\ }\href {\doibase 10.1007/JHEP12(2023)154} {\bibfield  {journal} {\bibinfo  {journal} {JHEP}\ }\textbf {\bibinfo {volume} {12}},\ \bibinfo {pages} {154} (\bibinfo {year} {2023})},\ \Eprint {http://arxiv.org/abs/2311.10130} {arXiv:2311.10130 [hep-ph]} \BibitemShut {NoStop}%
\bibitem [{\citenamefont {Boto}\ \emph {et~al.}(2024)\citenamefont {Boto}, \citenamefont {Das}, \citenamefont {Romao}, \citenamefont {Saha},\ and\ \citenamefont {Silva}}]{Boto:2023bpg}%
  \BibitemOpen
  \bibfield  {author} {\bibinfo {author} {\bibfnamefont {R.}~\bibnamefont {Boto}}, \bibinfo {author} {\bibfnamefont {D.}~\bibnamefont {Das}}, \bibinfo {author} {\bibfnamefont {J.~C.}\ \bibnamefont {Romao}}, \bibinfo {author} {\bibfnamefont {I.}~\bibnamefont {Saha}}, \ and\ \bibinfo {author} {\bibfnamefont {J.~P.}\ \bibnamefont {Silva}},\ }\href {\doibase 10.1103/PhysRevD.109.095002} {\bibfield  {journal} {\bibinfo  {journal} {Phys. Rev. D}\ }\textbf {\bibinfo {volume} {109}},\ \bibinfo {pages} {095002} (\bibinfo {year} {2024})},\ \Eprint {http://arxiv.org/abs/2312.13050} {arXiv:2312.13050 [hep-ph]} \BibitemShut {NoStop}%
\bibitem [{\citenamefont {Das}\ \emph {et~al.}(2024)\citenamefont {Das}, \citenamefont {Jha},\ and\ \citenamefont {Nanda}}]{Das:2024tfe}%
  \BibitemOpen
  \bibfield  {author} {\bibinfo {author} {\bibfnamefont {N.}~\bibnamefont {Das}}, \bibinfo {author} {\bibfnamefont {T.}~\bibnamefont {Jha}}, \ and\ \bibinfo {author} {\bibfnamefont {D.}~\bibnamefont {Nanda}},\ }\href {\doibase 10.1103/PhysRevD.109.115020} {\bibfield  {journal} {\bibinfo  {journal} {Phys. Rev. D}\ }\textbf {\bibinfo {volume} {109}},\ \bibinfo {pages} {115020} (\bibinfo {year} {2024})},\ \Eprint {http://arxiv.org/abs/2402.01317} {arXiv:2402.01317 [hep-ph]} \BibitemShut {NoStop}%
\bibitem [{\citenamefont {Cheung}\ and\ \citenamefont {Ouseph}(2024)}]{Cheung:2024kml}%
  \BibitemOpen
  \bibfield  {author} {\bibinfo {author} {\bibfnamefont {K.}~\bibnamefont {Cheung}}\ and\ \bibinfo {author} {\bibfnamefont {C.~J.}\ \bibnamefont {Ouseph}},\ }\href {\doibase 10.1103/PhysRevD.110.055016} {\bibfield  {journal} {\bibinfo  {journal} {Phys. Rev. D}\ }\textbf {\bibinfo {volume} {110}},\ \bibinfo {pages} {055016} (\bibinfo {year} {2024})},\ \Eprint {http://arxiv.org/abs/2402.05678} {arXiv:2402.05678 [hep-ph]} \BibitemShut {NoStop}%
\bibitem [{\citenamefont {Hu}\ \emph {et~al.}(2024)\citenamefont {Hu}, \citenamefont {Zhang}, \citenamefont {Zhu},\ and\ \citenamefont {Chen}}]{Hu:2024slu}%
  \BibitemOpen
  \bibfield  {author} {\bibinfo {author} {\bibfnamefont {H.-c.}\ \bibnamefont {Hu}}, \bibinfo {author} {\bibfnamefont {Z.-Y.}\ \bibnamefont {Zhang}}, \bibinfo {author} {\bibfnamefont {N.-Y.}\ \bibnamefont {Zhu}}, \ and\ \bibinfo {author} {\bibfnamefont {H.-X.}\ \bibnamefont {Chen}},\ }\href {\doibase 10.1088/1674-1137/ad5427} {\bibfield  {journal} {\bibinfo  {journal} {Chin. Phys. C}\ }\textbf {\bibinfo {volume} {48}},\ \bibinfo {pages} {093101} (\bibinfo {year} {2024})},\ \Eprint {http://arxiv.org/abs/2406.00946} {arXiv:2406.00946 [hep-ph]} \BibitemShut {NoStop}%
\bibitem [{\citenamefont {Hern\'andez-Ju\'arez}\ \emph {et~al.}(2024)\citenamefont {Hern\'andez-Ju\'arez}, \citenamefont {Gait\'an},\ and\ \citenamefont {Martinez}}]{Hernandez-Juarez:2024pty}%
  \BibitemOpen
  \bibfield  {author} {\bibinfo {author} {\bibfnamefont {A.~I.}\ \bibnamefont {Hern\'andez-Ju\'arez}}, \bibinfo {author} {\bibfnamefont {R.}~\bibnamefont {Gait\'an}}, \ and\ \bibinfo {author} {\bibfnamefont {R.}~\bibnamefont {Martinez}},\ }\href@noop {} {\  (\bibinfo {year} {2024})},\ \Eprint {http://arxiv.org/abs/2405.03094} {arXiv:2405.03094 [hep-ph]} \BibitemShut {NoStop}%
\bibitem [{\citenamefont {Zhang}\ \emph {et~al.}(2024)\citenamefont {Zhang}, \citenamefont {Yue}, \citenamefont {Sun},\ and\ \citenamefont {Wang}}]{Zhang:2024yqw}%
  \BibitemOpen
  \bibfield  {author} {\bibinfo {author} {\bibfnamefont {X.-M.}\ \bibnamefont {Zhang}}, \bibinfo {author} {\bibfnamefont {C.-X.}\ \bibnamefont {Yue}}, \bibinfo {author} {\bibfnamefont {X.-C.}\ \bibnamefont {Sun}}, \ and\ \bibinfo {author} {\bibfnamefont {Y.-Q.}\ \bibnamefont {Wang}},\ }\href {\doibase 10.1142/S0217751X24501124} {\bibfield  {journal} {\bibinfo  {journal} {Int. J. Mod. Phys. A}\ }\textbf {\bibinfo {volume} {39}},\ \bibinfo {pages} {2450112} (\bibinfo {year} {2024})}\BibitemShut {NoStop}%
\bibitem [{\citenamefont {Arhrib}\ \emph {et~al.}(2024)\citenamefont {Arhrib}, \citenamefont {Phan}, \citenamefont {Tran},\ and\ \citenamefont {Yuan}}]{Arhrib:2024wjj}%
  \BibitemOpen
  \bibfield  {author} {\bibinfo {author} {\bibfnamefont {A.}~\bibnamefont {Arhrib}}, \bibinfo {author} {\bibfnamefont {K.~H.}\ \bibnamefont {Phan}}, \bibinfo {author} {\bibfnamefont {V.~Q.}\ \bibnamefont {Tran}}, \ and\ \bibinfo {author} {\bibfnamefont {T.-C.}\ \bibnamefont {Yuan}},\ }\href@noop {} {\  (\bibinfo {year} {2024})},\ \Eprint {http://arxiv.org/abs/2405.03127} {arXiv:2405.03127 [hep-ph]} \BibitemShut {NoStop}%
\bibitem [{\citenamefont {Israr}\ and\ \citenamefont {Rehman}(2024)}]{Israr:2024ubp}%
  \BibitemOpen
  \bibfield  {author} {\bibinfo {author} {\bibfnamefont {S.}~\bibnamefont {Israr}}\ and\ \bibinfo {author} {\bibfnamefont {M.}~\bibnamefont {Rehman}},\ }\href@noop {} {\  (\bibinfo {year} {2024})},\ \Eprint {http://arxiv.org/abs/2407.01210} {arXiv:2407.01210 [hep-ph]} \BibitemShut {NoStop}%
\bibitem [{\citenamefont {Mantzaropoulos}(2024)}]{Mantzaropoulos:2024vpe}%
  \BibitemOpen
  \bibfield  {author} {\bibinfo {author} {\bibfnamefont {K.}~\bibnamefont {Mantzaropoulos}},\ }\href {\doibase 10.1103/PhysRevD.110.055041} {\bibfield  {journal} {\bibinfo  {journal} {Phys. Rev. D}\ }\textbf {\bibinfo {volume} {110}},\ \bibinfo {pages} {055041} (\bibinfo {year} {2024})},\ \Eprint {http://arxiv.org/abs/2407.09145} {arXiv:2407.09145 [hep-ph]} \BibitemShut {NoStop}%
\bibitem [{\citenamefont {Sang}\ \emph {et~al.}(2024)\citenamefont {Sang}, \citenamefont {Feng},\ and\ \citenamefont {Jia}}]{Sang:2024vqk}%
  \BibitemOpen
  \bibfield  {author} {\bibinfo {author} {\bibfnamefont {W.-L.}\ \bibnamefont {Sang}}, \bibinfo {author} {\bibfnamefont {F.}~\bibnamefont {Feng}}, \ and\ \bibinfo {author} {\bibfnamefont {Y.}~\bibnamefont {Jia}},\ }\href {\doibase 10.1103/PhysRevD.110.L051302} {\bibfield  {journal} {\bibinfo  {journal} {Phys. Rev. D}\ }\textbf {\bibinfo {volume} {110}},\ \bibinfo {pages} {L051302} (\bibinfo {year} {2024})},\ \Eprint {http://arxiv.org/abs/2405.03464} {arXiv:2405.03464 [hep-ph]} \BibitemShut {NoStop}%
\bibitem [{\citenamefont {Kachanovich}\ \emph {et~al.}(2020)\citenamefont {Kachanovich}, \citenamefont {Nierste},\ and\ \citenamefont {Ni\v{s}and\v{z}i\'c}}]{Kachanovich:2020xyg}%
  \BibitemOpen
  \bibfield  {author} {\bibinfo {author} {\bibfnamefont {A.}~\bibnamefont {Kachanovich}}, \bibinfo {author} {\bibfnamefont {U.}~\bibnamefont {Nierste}}, \ and\ \bibinfo {author} {\bibfnamefont {I.}~\bibnamefont {Ni\v{s}and\v{z}i\'c}},\ }\href {\doibase 10.1103/PhysRevD.101.073003} {\bibfield  {journal} {\bibinfo  {journal} {Phys. Rev. D}\ }\textbf {\bibinfo {volume} {101}},\ \bibinfo {pages} {073003} (\bibinfo {year} {2020})},\ \Eprint {http://arxiv.org/abs/2001.06516} {arXiv:2001.06516 [hep-ph]} \BibitemShut {NoStop}%
\bibitem [{\citenamefont {Kachanovich}\ \emph {et~al.}(2022)\citenamefont {Kachanovich}, \citenamefont {Nierste},\ and\ \citenamefont {Ni\v{s}and\v{z}i\'c}}]{Kachanovich:2021pvx}%
  \BibitemOpen
  \bibfield  {author} {\bibinfo {author} {\bibfnamefont {A.}~\bibnamefont {Kachanovich}}, \bibinfo {author} {\bibfnamefont {U.}~\bibnamefont {Nierste}}, \ and\ \bibinfo {author} {\bibfnamefont {I.}~\bibnamefont {Ni\v{s}and\v{z}i\'c}},\ }\href {\doibase 10.1103/PhysRevD.105.013007} {\bibfield  {journal} {\bibinfo  {journal} {Phys. Rev. D}\ }\textbf {\bibinfo {volume} {105}},\ \bibinfo {pages} {013007} (\bibinfo {year} {2022})},\ \Eprint {http://arxiv.org/abs/2109.04426} {arXiv:2109.04426 [hep-ph]} \BibitemShut {NoStop}%
\bibitem [{\citenamefont {Corbett}\ and\ \citenamefont {Rasmussen}(2022)}]{Corbett:2021iob}%
  \BibitemOpen
  \bibfield  {author} {\bibinfo {author} {\bibfnamefont {T.}~\bibnamefont {Corbett}}\ and\ \bibinfo {author} {\bibfnamefont {T.}~\bibnamefont {Rasmussen}},\ }\href {\doibase 10.21468/SciPostPhys.13.5.112} {\bibfield  {journal} {\bibinfo  {journal} {SciPost Phys.}\ }\textbf {\bibinfo {volume} {13}},\ \bibinfo {pages} {112} (\bibinfo {year} {2022})},\ \Eprint {http://arxiv.org/abs/2110.03694} {arXiv:2110.03694 [hep-ph]} \BibitemShut {NoStop}%
\bibitem [{\citenamefont {Chen}\ \emph {et~al.}(2022)\citenamefont {Chen}, \citenamefont {Gehrmann}, \citenamefont {Glover},\ and\ \citenamefont {Huss}}]{Chen:2021ibm}%
  \BibitemOpen
  \bibfield  {author} {\bibinfo {author} {\bibfnamefont {X.}~\bibnamefont {Chen}}, \bibinfo {author} {\bibfnamefont {T.}~\bibnamefont {Gehrmann}}, \bibinfo {author} {\bibfnamefont {E.~W.~N.}\ \bibnamefont {Glover}}, \ and\ \bibinfo {author} {\bibfnamefont {A.}~\bibnamefont {Huss}},\ }\href {\doibase 10.1007/JHEP01(2022)053} {\bibfield  {journal} {\bibinfo  {journal} {JHEP}\ }\textbf {\bibinfo {volume} {01}},\ \bibinfo {pages} {053} (\bibinfo {year} {2022})},\ \Eprint {http://arxiv.org/abs/2111.02157} {arXiv:2111.02157 [hep-ph]} \BibitemShut {NoStop}%
\bibitem [{\citenamefont {Ahmed}\ \emph {et~al.}(2024)\citenamefont {Ahmed}, \citenamefont {Hasan}, \citenamefont {Iqbal}, \citenamefont {Junaid}, \citenamefont {Tariq},\ and\ \citenamefont {Uzair}}]{Ahmed:2023vyl}%
  \BibitemOpen
  \bibfield  {author} {\bibinfo {author} {\bibfnamefont {I.}~\bibnamefont {Ahmed}}, \bibinfo {author} {\bibfnamefont {U.}~\bibnamefont {Hasan}}, \bibinfo {author} {\bibfnamefont {S.}~\bibnamefont {Iqbal}}, \bibinfo {author} {\bibfnamefont {M.}~\bibnamefont {Junaid}}, \bibinfo {author} {\bibfnamefont {B.}~\bibnamefont {Tariq}}, \ and\ \bibinfo {author} {\bibfnamefont {A.}~\bibnamefont {Uzair}},\ }\href {\doibase 10.1007/JHEP05(2024)187} {\bibfield  {journal} {\bibinfo  {journal} {JHEP}\ }\textbf {\bibinfo {volume} {05}},\ \bibinfo {pages} {187} (\bibinfo {year} {2024})},\ \Eprint {http://arxiv.org/abs/2309.07448} {arXiv:2309.07448 [hep-ph]} \BibitemShut {NoStop}%
\bibitem [{\citenamefont {Aakvaag}\ \emph {et~al.}(2024)\citenamefont {Aakvaag}, \citenamefont {Fomin}, \citenamefont {Lipniacka}, \citenamefont {Pokorski}, \citenamefont {Rosiek},\ and\ \citenamefont {Sahoo}}]{Aakvaag:2023xhy}%
  \BibitemOpen
  \bibfield  {author} {\bibinfo {author} {\bibfnamefont {E.}~\bibnamefont {Aakvaag}}, \bibinfo {author} {\bibfnamefont {N.}~\bibnamefont {Fomin}}, \bibinfo {author} {\bibfnamefont {A.}~\bibnamefont {Lipniacka}}, \bibinfo {author} {\bibfnamefont {S.}~\bibnamefont {Pokorski}}, \bibinfo {author} {\bibfnamefont {J.}~\bibnamefont {Rosiek}}, \ and\ \bibinfo {author} {\bibfnamefont {D.}~\bibnamefont {Sahoo}},\ }\href {\doibase 10.1140/epjc/s10052-024-12691-z} {\bibfield  {journal} {\bibinfo  {journal} {Eur. Phys. J. C}\ }\textbf {\bibinfo {volume} {84}},\ \bibinfo {pages} {341} (\bibinfo {year} {2024})},\ \Eprint {http://arxiv.org/abs/2311.16211} {arXiv:2311.16211 [hep-ph]} \BibitemShut {NoStop}%
\bibitem [{\citenamefont {Hue}\ \emph {et~al.}(2023)\citenamefont {Hue}, \citenamefont {Tran}, \citenamefont {Nguyen},\ and\ \citenamefont {Phan}}]{Hue:2023tdz}%
  \BibitemOpen
  \bibfield  {author} {\bibinfo {author} {\bibfnamefont {L.~T.}\ \bibnamefont {Hue}}, \bibinfo {author} {\bibfnamefont {D.~T.}\ \bibnamefont {Tran}}, \bibinfo {author} {\bibfnamefont {T.~H.}\ \bibnamefont {Nguyen}}, \ and\ \bibinfo {author} {\bibfnamefont {K.~H.}\ \bibnamefont {Phan}},\ }\href {\doibase 10.1093/ptep/ptad106} {\bibfield  {journal} {\bibinfo  {journal} {PTEP}\ }\textbf {\bibinfo {volume} {2023}},\ \bibinfo {pages} {083B06} (\bibinfo {year} {2023})},\ \Eprint {http://arxiv.org/abs/2305.04002} {arXiv:2305.04002 [hep-ph]} \BibitemShut {NoStop}%
\bibitem [{\citenamefont {Van~On}\ \emph {et~al.}(2022)\citenamefont {Van~On}, \citenamefont {Tran}, \citenamefont {Nguyen},\ and\ \citenamefont {Phan}}]{VanOn:2021myp}%
  \BibitemOpen
  \bibfield  {author} {\bibinfo {author} {\bibfnamefont {V.}~\bibnamefont {Van~On}}, \bibinfo {author} {\bibfnamefont {D.~T.}\ \bibnamefont {Tran}}, \bibinfo {author} {\bibfnamefont {C.~L.}\ \bibnamefont {Nguyen}}, \ and\ \bibinfo {author} {\bibfnamefont {K.~H.}\ \bibnamefont {Phan}},\ }\href {\doibase 10.1140/epjc/s10052-022-10225-z} {\bibfield  {journal} {\bibinfo  {journal} {Eur. Phys. J. C}\ }\textbf {\bibinfo {volume} {82}},\ \bibinfo {pages} {277} (\bibinfo {year} {2022})},\ \Eprint {http://arxiv.org/abs/2111.07708} {arXiv:2111.07708 [hep-ph]} \BibitemShut {NoStop}%
\bibitem [{\citenamefont {Sun}\ and\ \citenamefont {Gao}(2014)}]{Sun:2013cba}%
  \BibitemOpen
  \bibfield  {author} {\bibinfo {author} {\bibfnamefont {Y.}~\bibnamefont {Sun}}\ and\ \bibinfo {author} {\bibfnamefont {D.-N.}\ \bibnamefont {Gao}},\ }\href {\doibase 10.1103/PhysRevD.89.017301} {\bibfield  {journal} {\bibinfo  {journal} {Phys. Rev. D}\ }\textbf {\bibinfo {volume} {89}},\ \bibinfo {pages} {017301} (\bibinfo {year} {2014})},\ \Eprint {http://arxiv.org/abs/1310.8404} {arXiv:1310.8404 [hep-ph]} \BibitemShut {NoStop}%
\bibitem [{\citenamefont {Phan}\ \emph {et~al.}(2021)\citenamefont {Phan}, \citenamefont {Hue},\ and\ \citenamefont {Tran}}]{Phan:2021xwc}%
  \BibitemOpen
  \bibfield  {author} {\bibinfo {author} {\bibfnamefont {K.~H.}\ \bibnamefont {Phan}}, \bibinfo {author} {\bibfnamefont {L.~T.}\ \bibnamefont {Hue}}, \ and\ \bibinfo {author} {\bibfnamefont {D.~T.}\ \bibnamefont {Tran}},\ }\href {\doibase 10.1093/ptep/ptab121} {\bibfield  {journal} {\bibinfo  {journal} {PTEP}\ }\textbf {\bibinfo {volume} {2021}},\ \bibinfo {pages} {103B07} (\bibinfo {year} {2021})},\ \Eprint {http://arxiv.org/abs/2106.14466} {arXiv:2106.14466 [hep-ph]} \BibitemShut {NoStop}%
\bibitem [{\citenamefont {Phan}\ and\ \citenamefont {Tran}(2022)}]{Phan:2021ovj}%
  \BibitemOpen
  \bibfield  {author} {\bibinfo {author} {\bibfnamefont {K.~H.}\ \bibnamefont {Phan}}\ and\ \bibinfo {author} {\bibfnamefont {D.~T.}\ \bibnamefont {Tran}},\ }\href {\doibase 10.1093/ptep/ptac012} {\bibfield  {journal} {\bibinfo  {journal} {PTEP}\ }\textbf {\bibinfo {volume} {2022}},\ \bibinfo {pages} {023B03} (\bibinfo {year} {2022})},\ \Eprint {http://arxiv.org/abs/2111.07698} {arXiv:2111.07698 [hep-ph]} \BibitemShut {NoStop}%
\bibitem [{\citenamefont {Kachanovich}\ and\ \citenamefont {Ni\v{s}and\v{z}i\'c}(2024)}]{Kachanovich:2024vpt}%
  \BibitemOpen
  \bibfield  {author} {\bibinfo {author} {\bibfnamefont {A.}~\bibnamefont {Kachanovich}}\ and\ \bibinfo {author} {\bibfnamefont {I.}~\bibnamefont {Ni\v{s}and\v{z}i\'c}},\ }\href@noop {} {\  (\bibinfo {year} {2024})},\ \Eprint {http://arxiv.org/abs/2405.16239} {arXiv:2405.16239 [hep-ph]} \BibitemShut {NoStop}%
\bibitem [{\citenamefont {Abbasabadi}\ \emph {et~al.}(1997)\citenamefont {Abbasabadi}, \citenamefont {Bowser-Chao}, \citenamefont {Dicus},\ and\ \citenamefont {Repko}}]{Abbasabadi:1996ze}%
  \BibitemOpen
  \bibfield  {author} {\bibinfo {author} {\bibfnamefont {A.}~\bibnamefont {Abbasabadi}}, \bibinfo {author} {\bibfnamefont {D.}~\bibnamefont {Bowser-Chao}}, \bibinfo {author} {\bibfnamefont {D.~A.}\ \bibnamefont {Dicus}}, \ and\ \bibinfo {author} {\bibfnamefont {W.~W.}\ \bibnamefont {Repko}},\ }\href {\doibase 10.1103/PhysRevD.55.5647} {\bibfield  {journal} {\bibinfo  {journal} {Phys. Rev. D}\ }\textbf {\bibinfo {volume} {55}},\ \bibinfo {pages} {5647} (\bibinfo {year} {1997})},\ \Eprint {http://arxiv.org/abs/hep-ph/9611209} {arXiv:hep-ph/9611209} \BibitemShut {NoStop}%
\bibitem [{\citenamefont {Chen}\ \emph {et~al.}(2013)\citenamefont {Chen}, \citenamefont {Qiao},\ and\ \citenamefont {Zhu}}]{Chen:2012ju}%
  \BibitemOpen
  \bibfield  {author} {\bibinfo {author} {\bibfnamefont {L.-B.}\ \bibnamefont {Chen}}, \bibinfo {author} {\bibfnamefont {C.-F.}\ \bibnamefont {Qiao}}, \ and\ \bibinfo {author} {\bibfnamefont {R.-L.}\ \bibnamefont {Zhu}},\ }\href {\doibase 10.1016/j.physletb.2013.08.050} {\bibfield  {journal} {\bibinfo  {journal} {Phys. Lett. B}\ }\textbf {\bibinfo {volume} {726}},\ \bibinfo {pages} {306} (\bibinfo {year} {2013})},\ \bibinfo {note} {[Erratum: Phys.Lett.B 808, 135629 (2020)]},\ \Eprint {http://arxiv.org/abs/1211.6058} {arXiv:1211.6058 [hep-ph]} \BibitemShut {NoStop}%
\bibitem [{\citenamefont {Dicus}\ and\ \citenamefont {Repko}(2013)}]{Dicus:2013ycd}%
  \BibitemOpen
  \bibfield  {author} {\bibinfo {author} {\bibfnamefont {D.~A.}\ \bibnamefont {Dicus}}\ and\ \bibinfo {author} {\bibfnamefont {W.~W.}\ \bibnamefont {Repko}},\ }\href {\doibase 10.1103/PhysRevD.87.077301} {\bibfield  {journal} {\bibinfo  {journal} {Phys. Rev. D}\ }\textbf {\bibinfo {volume} {87}},\ \bibinfo {pages} {077301} (\bibinfo {year} {2013})},\ \Eprint {http://arxiv.org/abs/1302.2159} {arXiv:1302.2159 [hep-ph]} \BibitemShut {NoStop}%
\bibitem [{\citenamefont {Passarino}(2013)}]{Passarino:2013nka}%
  \BibitemOpen
  \bibfield  {author} {\bibinfo {author} {\bibfnamefont {G.}~\bibnamefont {Passarino}},\ }\href {\doibase 10.1016/j.physletb.2013.10.052} {\bibfield  {journal} {\bibinfo  {journal} {Phys. Lett. B}\ }\textbf {\bibinfo {volume} {727}},\ \bibinfo {pages} {424} (\bibinfo {year} {2013})},\ \Eprint {http://arxiv.org/abs/1308.0422} {arXiv:1308.0422 [hep-ph]} \BibitemShut {NoStop}%
\bibitem [{\citenamefont {Han}\ and\ \citenamefont {Wang}(2017)}]{Han:2017yhy}%
  \BibitemOpen
  \bibfield  {author} {\bibinfo {author} {\bibfnamefont {T.}~\bibnamefont {Han}}\ and\ \bibinfo {author} {\bibfnamefont {X.}~\bibnamefont {Wang}},\ }\href {\doibase 10.1007/JHEP10(2017)036} {\bibfield  {journal} {\bibinfo  {journal} {JHEP}\ }\textbf {\bibinfo {volume} {10}},\ \bibinfo {pages} {036} (\bibinfo {year} {2017})},\ \Eprint {http://arxiv.org/abs/1704.00790} {arXiv:1704.00790 [hep-ph]} \BibitemShut {NoStop}%
\bibitem [{\citenamefont {Buccioni}\ \emph {et~al.}(2024)\citenamefont {Buccioni}, \citenamefont {Devoto}, \citenamefont {Djouadi}, \citenamefont {Ellis}, \citenamefont {Quevillon},\ and\ \citenamefont {Tancredi}}]{Buccioni:2023qnt}%
  \BibitemOpen
  \bibfield  {author} {\bibinfo {author} {\bibfnamefont {F.}~\bibnamefont {Buccioni}}, \bibinfo {author} {\bibfnamefont {F.}~\bibnamefont {Devoto}}, \bibinfo {author} {\bibfnamefont {A.}~\bibnamefont {Djouadi}}, \bibinfo {author} {\bibfnamefont {J.}~\bibnamefont {Ellis}}, \bibinfo {author} {\bibfnamefont {J.}~\bibnamefont {Quevillon}}, \ and\ \bibinfo {author} {\bibfnamefont {L.}~\bibnamefont {Tancredi}},\ }\href {\doibase 10.1016/j.physletb.2024.138596} {\bibfield  {journal} {\bibinfo  {journal} {Phys. Lett. B}\ }\textbf {\bibinfo {volume} {851}},\ \bibinfo {pages} {138596} (\bibinfo {year} {2024})},\ \Eprint {http://arxiv.org/abs/2312.12384} {arXiv:2312.12384 [hep-ph]} \BibitemShut {NoStop}%
\bibitem [{\citenamefont {Chen}\ \emph {et~al.}(2024)\citenamefont {Chen}, \citenamefont {Chen}, \citenamefont {Qiao},\ and\ \citenamefont {Zhu}}]{Chen:2024vyn}%
  \BibitemOpen
  \bibfield  {author} {\bibinfo {author} {\bibfnamefont {Z.-Q.}\ \bibnamefont {Chen}}, \bibinfo {author} {\bibfnamefont {L.-B.}\ \bibnamefont {Chen}}, \bibinfo {author} {\bibfnamefont {C.-F.}\ \bibnamefont {Qiao}}, \ and\ \bibinfo {author} {\bibfnamefont {R.}~\bibnamefont {Zhu}},\ }\href {\doibase 10.1103/PhysRevD.110.L051301} {\bibfield  {journal} {\bibinfo  {journal} {Phys. Rev. D}\ }\textbf {\bibinfo {volume} {110}},\ \bibinfo {pages} {L051301} (\bibinfo {year} {2024})},\ \Eprint {http://arxiv.org/abs/2404.11441} {arXiv:2404.11441 [hep-ph]} \BibitemShut {NoStop}%
\bibitem [{\citenamefont {Alonso-\'Alvarez}\ \emph {et~al.}(2023)\citenamefont {Alonso-\'Alvarez}, \citenamefont {Jaeckel},\ and\ \citenamefont {Lopes}}]{Alonso-Alvarez:2023wni}%
  \BibitemOpen
  \bibfield  {author} {\bibinfo {author} {\bibfnamefont {G.}~\bibnamefont {Alonso-\'Alvarez}}, \bibinfo {author} {\bibfnamefont {J.}~\bibnamefont {Jaeckel}}, \ and\ \bibinfo {author} {\bibfnamefont {D.~D.}\ \bibnamefont {Lopes}},\ }\href@noop {} {\  (\bibinfo {year} {2023})},\ \Eprint {http://arxiv.org/abs/2302.12262} {arXiv:2302.12262 [hep-ph]} \BibitemShut {NoStop}%
\bibitem [{\citenamefont {Peskin}\ and\ \citenamefont {Schroeder}(1995)}]{Peskin:1995ev}%
  \BibitemOpen
  \bibfield  {author} {\bibinfo {author} {\bibfnamefont {M.~E.}\ \bibnamefont {Peskin}}\ and\ \bibinfo {author} {\bibfnamefont {D.~V.}\ \bibnamefont {Schroeder}},\ }\href {\doibase 10.1201/9780429503559} {\emph {\bibinfo {title} {{An Introduction to quantum field theory}}}}\ (\bibinfo  {publisher} {Addison-Wesley},\ \bibinfo {address} {Reading, USA},\ \bibinfo {year} {1995})\BibitemShut {NoStop}%
\bibitem [{\citenamefont {Grzadkowski}\ \emph {et~al.}(2010)\citenamefont {Grzadkowski}, \citenamefont {Iskrzynski}, \citenamefont {Misiak},\ and\ \citenamefont {Rosiek}}]{Grzadkowski:2010es}%
  \BibitemOpen
  \bibfield  {author} {\bibinfo {author} {\bibfnamefont {B.}~\bibnamefont {Grzadkowski}}, \bibinfo {author} {\bibfnamefont {M.}~\bibnamefont {Iskrzynski}}, \bibinfo {author} {\bibfnamefont {M.}~\bibnamefont {Misiak}}, \ and\ \bibinfo {author} {\bibfnamefont {J.}~\bibnamefont {Rosiek}},\ }\href {\doibase 10.1007/JHEP10(2010)085} {\bibfield  {journal} {\bibinfo  {journal} {JHEP}\ }\textbf {\bibinfo {volume} {10}},\ \bibinfo {pages} {085} (\bibinfo {year} {2010})},\ \Eprint {http://arxiv.org/abs/1008.4884} {arXiv:1008.4884 [hep-ph]} \BibitemShut {NoStop}%
\bibitem [{\citenamefont {Barger}\ \emph {et~al.}(2009)\citenamefont {Barger}, \citenamefont {Gao}, \citenamefont {Keung},\ and\ \citenamefont {Marfatia}}]{Barger:2009xe}%
  \BibitemOpen
  \bibfield  {author} {\bibinfo {author} {\bibfnamefont {V.}~\bibnamefont {Barger}}, \bibinfo {author} {\bibfnamefont {Y.}~\bibnamefont {Gao}}, \bibinfo {author} {\bibfnamefont {W.~Y.}\ \bibnamefont {Keung}}, \ and\ \bibinfo {author} {\bibfnamefont {D.}~\bibnamefont {Marfatia}},\ }\href {\doibase 10.1103/PhysRevD.80.063537} {\bibfield  {journal} {\bibinfo  {journal} {Phys. Rev. D}\ }\textbf {\bibinfo {volume} {80}},\ \bibinfo {pages} {063537} (\bibinfo {year} {2009})},\ \Eprint {http://arxiv.org/abs/0906.3009} {arXiv:0906.3009 [hep-ph]} \BibitemShut {NoStop}%
\bibitem [{\citenamefont {Bergstrom}\ and\ \citenamefont {Snellman}(1988)}]{Bergstrom:1988fp}%
  \BibitemOpen
  \bibfield  {author} {\bibinfo {author} {\bibfnamefont {L.}~\bibnamefont {Bergstrom}}\ and\ \bibinfo {author} {\bibfnamefont {H.}~\bibnamefont {Snellman}},\ }\href {\doibase 10.1103/PhysRevD.37.3737} {\bibfield  {journal} {\bibinfo  {journal} {Phys. Rev. D}\ }\textbf {\bibinfo {volume} {37}},\ \bibinfo {pages} {3737} (\bibinfo {year} {1988})}\BibitemShut {NoStop}%
\bibitem [{\citenamefont {Bringmann}\ \emph {et~al.}(2012)\citenamefont {Bringmann}, \citenamefont {Huang}, \citenamefont {Ibarra}, \citenamefont {Vogl},\ and\ \citenamefont {Weniger}}]{Bringmann:2012vr}%
  \BibitemOpen
  \bibfield  {author} {\bibinfo {author} {\bibfnamefont {T.}~\bibnamefont {Bringmann}}, \bibinfo {author} {\bibfnamefont {X.}~\bibnamefont {Huang}}, \bibinfo {author} {\bibfnamefont {A.}~\bibnamefont {Ibarra}}, \bibinfo {author} {\bibfnamefont {S.}~\bibnamefont {Vogl}}, \ and\ \bibinfo {author} {\bibfnamefont {C.}~\bibnamefont {Weniger}},\ }\href {\doibase 10.1088/1475-7516/2012/07/054} {\bibfield  {journal} {\bibinfo  {journal} {JCAP}\ }\textbf {\bibinfo {volume} {07}},\ \bibinfo {pages} {054} (\bibinfo {year} {2012})},\ \Eprint {http://arxiv.org/abs/1203.1312} {arXiv:1203.1312 [hep-ph]} \BibitemShut {NoStop}%
\bibitem [{\citenamefont {Toma}(2013)}]{Toma:2013bka}%
  \BibitemOpen
  \bibfield  {author} {\bibinfo {author} {\bibfnamefont {T.}~\bibnamefont {Toma}},\ }\href {\doibase 10.1103/PhysRevLett.111.091301} {\bibfield  {journal} {\bibinfo  {journal} {Phys. Rev. Lett.}\ }\textbf {\bibinfo {volume} {111}},\ \bibinfo {pages} {091301} (\bibinfo {year} {2013})},\ \Eprint {http://arxiv.org/abs/1307.6181} {arXiv:1307.6181 [hep-ph]} \BibitemShut {NoStop}%
\bibitem [{\citenamefont {Giacchino}\ \emph {et~al.}(2013)\citenamefont {Giacchino}, \citenamefont {Lopez-Honorez},\ and\ \citenamefont {Tytgat}}]{Giacchino:2013bta}%
  \BibitemOpen
  \bibfield  {author} {\bibinfo {author} {\bibfnamefont {F.}~\bibnamefont {Giacchino}}, \bibinfo {author} {\bibfnamefont {L.}~\bibnamefont {Lopez-Honorez}}, \ and\ \bibinfo {author} {\bibfnamefont {M.~H.~G.}\ \bibnamefont {Tytgat}},\ }\href {\doibase 10.1088/1475-7516/2013/10/025} {\bibfield  {journal} {\bibinfo  {journal} {JCAP}\ }\textbf {\bibinfo {volume} {10}},\ \bibinfo {pages} {025} (\bibinfo {year} {2013})},\ \Eprint {http://arxiv.org/abs/1307.6480} {arXiv:1307.6480 [hep-ph]} \BibitemShut {NoStop}%
\bibitem [{\citenamefont {Giacchino}\ \emph {et~al.}(2014)\citenamefont {Giacchino}, \citenamefont {Lopez-Honorez},\ and\ \citenamefont {Tytgat}}]{Giacchino:2014moa}%
  \BibitemOpen
  \bibfield  {author} {\bibinfo {author} {\bibfnamefont {F.}~\bibnamefont {Giacchino}}, \bibinfo {author} {\bibfnamefont {L.}~\bibnamefont {Lopez-Honorez}}, \ and\ \bibinfo {author} {\bibfnamefont {M.~H.~G.}\ \bibnamefont {Tytgat}},\ }\href {\doibase 10.1088/1475-7516/2014/08/046} {\bibfield  {journal} {\bibinfo  {journal} {JCAP}\ }\textbf {\bibinfo {volume} {08}},\ \bibinfo {pages} {046} (\bibinfo {year} {2014})},\ \Eprint {http://arxiv.org/abs/1405.6921} {arXiv:1405.6921 [hep-ph]} \BibitemShut {NoStop}%
\bibitem [{\citenamefont {Silveira}\ and\ \citenamefont {Zee}(1985)}]{Silveira:1985rk}%
  \BibitemOpen
  \bibfield  {author} {\bibinfo {author} {\bibfnamefont {V.}~\bibnamefont {Silveira}}\ and\ \bibinfo {author} {\bibfnamefont {A.}~\bibnamefont {Zee}},\ }\href {\doibase 10.1016/0370-2693(85)90624-0} {\bibfield  {journal} {\bibinfo  {journal} {Phys. Lett. B}\ }\textbf {\bibinfo {volume} {161}},\ \bibinfo {pages} {136} (\bibinfo {year} {1985})}\BibitemShut {NoStop}%
\bibitem [{\citenamefont {McDonald}(1994)}]{McDonald:1993ex}%
  \BibitemOpen
  \bibfield  {author} {\bibinfo {author} {\bibfnamefont {J.}~\bibnamefont {McDonald}},\ }\href {\doibase 10.1103/PhysRevD.50.3637} {\bibfield  {journal} {\bibinfo  {journal} {Phys. Rev. D}\ }\textbf {\bibinfo {volume} {50}},\ \bibinfo {pages} {3637} (\bibinfo {year} {1994})},\ \Eprint {http://arxiv.org/abs/hep-ph/0702143} {arXiv:hep-ph/0702143} \BibitemShut {NoStop}%
\bibitem [{\citenamefont {Burgess}\ \emph {et~al.}(2001)\citenamefont {Burgess}, \citenamefont {Pospelov},\ and\ \citenamefont {ter Veldhuis}}]{Burgess:2000yq}%
  \BibitemOpen
  \bibfield  {author} {\bibinfo {author} {\bibfnamefont {C.~P.}\ \bibnamefont {Burgess}}, \bibinfo {author} {\bibfnamefont {M.}~\bibnamefont {Pospelov}}, \ and\ \bibinfo {author} {\bibfnamefont {T.}~\bibnamefont {ter Veldhuis}},\ }\href {\doibase 10.1016/S0550-3213(01)00513-2} {\bibfield  {journal} {\bibinfo  {journal} {Nucl. Phys. B}\ }\textbf {\bibinfo {volume} {619}},\ \bibinfo {pages} {709} (\bibinfo {year} {2001})},\ \Eprint {http://arxiv.org/abs/hep-ph/0011335} {arXiv:hep-ph/0011335} \BibitemShut {NoStop}%
\bibitem [{\citenamefont {Patt}\ and\ \citenamefont {Wilczek}(2006)}]{Patt:2006fw}%
  \BibitemOpen
  \bibfield  {author} {\bibinfo {author} {\bibfnamefont {B.}~\bibnamefont {Patt}}\ and\ \bibinfo {author} {\bibfnamefont {F.}~\bibnamefont {Wilczek}},\ }\href@noop {} {\  (\bibinfo {year} {2006})},\ \Eprint {http://arxiv.org/abs/hep-ph/0605188} {arXiv:hep-ph/0605188} \BibitemShut {NoStop}%
\bibitem [{\citenamefont {Navas}\ \emph {et~al.}(2024)\citenamefont {Navas} \emph {et~al.}}]{ParticleDataGroup:2024cfk}%
  \BibitemOpen
  \bibfield  {author} {\bibinfo {author} {\bibfnamefont {S.}~\bibnamefont {Navas}} \emph {et~al.} (\bibinfo {collaboration} {Particle Data Group}),\ }\href {\doibase 10.1103/PhysRevD.110.030001} {\bibfield  {journal} {\bibinfo  {journal} {Phys. Rev. D}\ }\textbf {\bibinfo {volume} {110}},\ \bibinfo {pages} {030001} (\bibinfo {year} {2024})}\BibitemShut {NoStop}%
\bibitem [{\citenamefont {Szabo}\ \emph {et~al.}(2024)\citenamefont {Szabo}, \citenamefont {Lellouch}, \citenamefont {Fodor}, \citenamefont {Stokes}, \citenamefont {Toth},\ and\ \citenamefont {Wang}}]{Szabo:2024pjz}%
  \BibitemOpen
  \bibfield  {author} {\bibinfo {author} {\bibfnamefont {K.~K.}\ \bibnamefont {Szabo}}, \bibinfo {author} {\bibfnamefont {L.}~\bibnamefont {Lellouch}}, \bibinfo {author} {\bibfnamefont {Z.}~\bibnamefont {Fodor}}, \bibinfo {author} {\bibfnamefont {F.}~\bibnamefont {Stokes}}, \bibinfo {author} {\bibfnamefont {B.~C.}\ \bibnamefont {Toth}}, \ and\ \bibinfo {author} {\bibfnamefont {G.}~\bibnamefont {Wang}},\ }\href {\doibase 10.1016/j.procs.2024.07.012} {\bibfield  {journal} {\bibinfo  {journal} {Procedia Comput. Sci.}\ }\textbf {\bibinfo {volume} {240}},\ \bibinfo {pages} {91} (\bibinfo {year} {2024})}\BibitemShut {NoStop}%
\bibitem [{\citenamefont {Abi}\ \emph {et~al.}(2021)\citenamefont {Abi} \emph {et~al.}}]{Muong-2:2021ojo}%
  \BibitemOpen
  \bibfield  {author} {\bibinfo {author} {\bibfnamefont {B.}~\bibnamefont {Abi}} \emph {et~al.} (\bibinfo {collaboration} {Muon g-2}),\ }\href {\doibase 10.1103/PhysRevLett.126.141801} {\bibfield  {journal} {\bibinfo  {journal} {Phys. Rev. Lett.}\ }\textbf {\bibinfo {volume} {126}},\ \bibinfo {pages} {141801} (\bibinfo {year} {2021})},\ \Eprint {http://arxiv.org/abs/2104.03281} {arXiv:2104.03281 [hep-ex]} \BibitemShut {NoStop}%
\bibitem [{\citenamefont {Peskin}\ and\ \citenamefont {Takeuchi}(1990)}]{Peskin:1990zt}%
  \BibitemOpen
  \bibfield  {author} {\bibinfo {author} {\bibfnamefont {M.~E.}\ \bibnamefont {Peskin}}\ and\ \bibinfo {author} {\bibfnamefont {T.}~\bibnamefont {Takeuchi}},\ }\href {\doibase 10.1103/PhysRevLett.65.964} {\bibfield  {journal} {\bibinfo  {journal} {Phys. Rev. Lett.}\ }\textbf {\bibinfo {volume} {65}},\ \bibinfo {pages} {964} (\bibinfo {year} {1990})}\BibitemShut {NoStop}%
\bibitem [{\citenamefont {Maksymyk}\ \emph {et~al.}(1994)\citenamefont {Maksymyk}, \citenamefont {Burgess},\ and\ \citenamefont {London}}]{Maksymyk:1993zm}%
  \BibitemOpen
  \bibfield  {author} {\bibinfo {author} {\bibfnamefont {I.}~\bibnamefont {Maksymyk}}, \bibinfo {author} {\bibfnamefont {C.~P.}\ \bibnamefont {Burgess}}, \ and\ \bibinfo {author} {\bibfnamefont {D.}~\bibnamefont {London}},\ }\href {\doibase 10.1103/PhysRevD.50.529} {\bibfield  {journal} {\bibinfo  {journal} {Phys. Rev. D}\ }\textbf {\bibinfo {volume} {50}},\ \bibinfo {pages} {529} (\bibinfo {year} {1994})},\ \Eprint {http://arxiv.org/abs/hep-ph/9306267} {arXiv:hep-ph/9306267} \BibitemShut {NoStop}%
\bibitem [{\citenamefont {Albergaria}\ \emph {et~al.}(2024)\citenamefont {Albergaria}, \citenamefont {Jur\v{c}iukonis},\ and\ \citenamefont {Lavoura}}]{Albergaria:2023nby}%
  \BibitemOpen
  \bibfield  {author} {\bibinfo {author} {\bibfnamefont {F.}~\bibnamefont {Albergaria}}, \bibinfo {author} {\bibfnamefont {D.}~\bibnamefont {Jur\v{c}iukonis}}, \ and\ \bibinfo {author} {\bibfnamefont {L.}~\bibnamefont {Lavoura}},\ }\href {\doibase 10.1007/JHEP05(2024)190} {\bibfield  {journal} {\bibinfo  {journal} {JHEP}\ }\textbf {\bibinfo {volume} {05}},\ \bibinfo {pages} {190} (\bibinfo {year} {2024})},\ \Eprint {http://arxiv.org/abs/2312.09099} {arXiv:2312.09099 [hep-ph]} \BibitemShut {NoStop}%
\bibitem [{\citenamefont {Moreno}\ and\ \citenamefont {Tytgat}(1992)}]{Moreno:1991mf}%
  \BibitemOpen
  \bibfield  {author} {\bibinfo {author} {\bibfnamefont {J.~M.}\ \bibnamefont {Moreno}}\ and\ \bibinfo {author} {\bibfnamefont {M.}~\bibnamefont {Tytgat}},\ }\href {\doibase 10.1007/BF01558303} {\bibfield  {journal} {\bibinfo  {journal} {Z. Phys. C}\ }\textbf {\bibinfo {volume} {55}},\ \bibinfo {pages} {175} (\bibinfo {year} {1992})}\BibitemShut {NoStop}%
\bibitem [{\citenamefont {Collaboration}(2024)}]{cmscollaboration2024darksectorsearchescms}%
  \BibitemOpen
  \bibfield  {author} {\bibinfo {author} {\bibfnamefont {C.}~\bibnamefont {Collaboration}},\ }\href {https://arxiv.org/abs/2405.13778} {\enquote {\bibinfo {title} {Dark sector searches with the cms experiment},}\ } (\bibinfo {year} {2024}),\ \Eprint {http://arxiv.org/abs/2405.13778} {arXiv:2405.13778 [hep-ex]} \BibitemShut {NoStop}%
\bibitem [{\citenamefont {Arcadi}\ \emph {et~al.}(2024)\citenamefont {Arcadi}, \citenamefont {Costa}, \citenamefont {Goudelis},\ and\ \citenamefont {Lebedev}}]{Arcadi:2024wwg}%
  \BibitemOpen
  \bibfield  {author} {\bibinfo {author} {\bibfnamefont {G.}~\bibnamefont {Arcadi}}, \bibinfo {author} {\bibfnamefont {F.}~\bibnamefont {Costa}}, \bibinfo {author} {\bibfnamefont {A.}~\bibnamefont {Goudelis}}, \ and\ \bibinfo {author} {\bibfnamefont {O.}~\bibnamefont {Lebedev}},\ }\href {\doibase 10.1007/JHEP07(2024)044} {\bibfield  {journal} {\bibinfo  {journal} {JHEP}\ }\textbf {\bibinfo {volume} {07}},\ \bibinfo {pages} {044} (\bibinfo {year} {2024})},\ \Eprint {http://arxiv.org/abs/2405.03760} {arXiv:2405.03760 [hep-ph]} \BibitemShut {NoStop}%
\bibitem [{\citenamefont {Arina}\ \emph {et~al.}(2023)\citenamefont {Arina}, \citenamefont {Fuks}, \citenamefont {Heisig}, \citenamefont {Kr\"amer}, \citenamefont {Mantani},\ and\ \citenamefont {Panizzi}}]{Arina:2023msd}%
  \BibitemOpen
  \bibfield  {author} {\bibinfo {author} {\bibfnamefont {C.}~\bibnamefont {Arina}}, \bibinfo {author} {\bibfnamefont {B.}~\bibnamefont {Fuks}}, \bibinfo {author} {\bibfnamefont {J.}~\bibnamefont {Heisig}}, \bibinfo {author} {\bibfnamefont {M.}~\bibnamefont {Kr\"amer}}, \bibinfo {author} {\bibfnamefont {L.}~\bibnamefont {Mantani}}, \ and\ \bibinfo {author} {\bibfnamefont {L.}~\bibnamefont {Panizzi}},\ }\href {\doibase 10.1103/PhysRevD.108.115007} {\bibfield  {journal} {\bibinfo  {journal} {Phys. Rev. D}\ }\textbf {\bibinfo {volume} {108}},\ \bibinfo {pages} {115007} (\bibinfo {year} {2023})},\ \Eprint {http://arxiv.org/abs/2307.10367} {arXiv:2307.10367 [hep-ph]} \BibitemShut {NoStop}%
\bibitem [{\citenamefont {Kachanovich}(2021)}]{Kachanovich:2020dah}%
  \BibitemOpen
  \bibfield  {author} {\bibinfo {author} {\bibfnamefont {A.}~\bibnamefont {Kachanovich}},\ }\emph {\bibinfo {title} {{Flavour-changing neutral current processes beyond the Standard Model}}},\ \href {\doibase 10.5445/IR/1000130547} {Ph.D. thesis},\ \bibinfo  {school} {KIT, Karlsruhe, Dept. Phys.} (\bibinfo {year} {2021})\BibitemShut {NoStop}%
\bibitem [{\citenamefont {Denner}\ \emph {et~al.}(2017)\citenamefont {Denner}, \citenamefont {Dittmaier},\ and\ \citenamefont {Hofer}}]{Denner:2016kdg}%
  \BibitemOpen
  \bibfield  {author} {\bibinfo {author} {\bibfnamefont {A.}~\bibnamefont {Denner}}, \bibinfo {author} {\bibfnamefont {S.}~\bibnamefont {Dittmaier}}, \ and\ \bibinfo {author} {\bibfnamefont {L.}~\bibnamefont {Hofer}},\ }\href {\doibase 10.1016/j.cpc.2016.10.013} {\bibfield  {journal} {\bibinfo  {journal} {Comput. Phys. Commun.}\ }\textbf {\bibinfo {volume} {212}},\ \bibinfo {pages} {220} (\bibinfo {year} {2017})},\ \Eprint {http://arxiv.org/abs/1604.06792} {arXiv:1604.06792 [hep-ph]} \BibitemShut {NoStop}%
\bibitem [{\citenamefont {Denner}\ and\ \citenamefont {Dittmaier}(2006)}]{Denner:2005nn}%
  \BibitemOpen
  \bibfield  {author} {\bibinfo {author} {\bibfnamefont {A.}~\bibnamefont {Denner}}\ and\ \bibinfo {author} {\bibfnamefont {S.}~\bibnamefont {Dittmaier}},\ }\href {\doibase 10.1016/j.nuclphysb.2005.11.007} {\bibfield  {journal} {\bibinfo  {journal} {Nucl. Phys. B}\ }\textbf {\bibinfo {volume} {734}},\ \bibinfo {pages} {62} (\bibinfo {year} {2006})},\ \Eprint {http://arxiv.org/abs/hep-ph/0509141} {arXiv:hep-ph/0509141} \BibitemShut {NoStop}%
\bibitem [{\citenamefont {Denner}\ and\ \citenamefont {Dittmaier}(2011)}]{Denner:2010tr}%
  \BibitemOpen
  \bibfield  {author} {\bibinfo {author} {\bibfnamefont {A.}~\bibnamefont {Denner}}\ and\ \bibinfo {author} {\bibfnamefont {S.}~\bibnamefont {Dittmaier}},\ }\href {\doibase 10.1016/j.nuclphysb.2010.11.002} {\bibfield  {journal} {\bibinfo  {journal} {Nucl. Phys. B}\ }\textbf {\bibinfo {volume} {844}},\ \bibinfo {pages} {199} (\bibinfo {year} {2011})},\ \Eprint {http://arxiv.org/abs/1005.2076} {arXiv:1005.2076 [hep-ph]} \BibitemShut {NoStop}%
\bibitem [{\citenamefont {Inc.}()}]{Mathematica}%
  \BibitemOpen
  \bibfield  {author} {\bibinfo {author} {\bibfnamefont {W.~R.}\ \bibnamefont {Inc.}},\ }\href@noop {} {\enquote {\bibinfo {title} {Mathematica, {V}ersion 12.0},}\ }\bibinfo {note} {Champaign, IL, 2019}\BibitemShut {NoStop}%
\bibitem [{\citenamefont {Patel}(2015)}]{Patel:2015tea}%
  \BibitemOpen
  \bibfield  {author} {\bibinfo {author} {\bibfnamefont {H.~H.}\ \bibnamefont {Patel}},\ }\href {\doibase 10.1016/j.cpc.2015.08.017} {\bibfield  {journal} {\bibinfo  {journal} {Comput. Phys. Commun.}\ }\textbf {\bibinfo {volume} {197}},\ \bibinfo {pages} {276} (\bibinfo {year} {2015})},\ \Eprint {http://arxiv.org/abs/1503.01469} {arXiv:1503.01469 [hep-ph]} \BibitemShut {NoStop}%
\bibitem [{\citenamefont {Patel}(2017)}]{Patel:2016fam}%
  \BibitemOpen
  \bibfield  {author} {\bibinfo {author} {\bibfnamefont {H.~H.}\ \bibnamefont {Patel}},\ }\href {\doibase 10.1016/j.cpc.2017.04.015} {\bibfield  {journal} {\bibinfo  {journal} {Comput. Phys. Commun.}\ }\textbf {\bibinfo {volume} {218}},\ \bibinfo {pages} {66} (\bibinfo {year} {2017})},\ \Eprint {http://arxiv.org/abs/1612.00009} {arXiv:1612.00009 [hep-ph]} \BibitemShut {NoStop}%
\bibitem [{\citenamefont {Hahn}\ and\ \citenamefont {Perez-Victoria}(1999)}]{Hahn:1998yk}%
  \BibitemOpen
  \bibfield  {author} {\bibinfo {author} {\bibfnamefont {T.}~\bibnamefont {Hahn}}\ and\ \bibinfo {author} {\bibfnamefont {M.}~\bibnamefont {Perez-Victoria}},\ }\href {\doibase 10.1016/S0010-4655(98)00173-8} {\bibfield  {journal} {\bibinfo  {journal} {Comput. Phys. Commun.}\ }\textbf {\bibinfo {volume} {118}},\ \bibinfo {pages} {153} (\bibinfo {year} {1999})},\ \Eprint {http://arxiv.org/abs/hep-ph/9807565} {arXiv:hep-ph/9807565} \BibitemShut {NoStop}%
\bibitem [{\citenamefont {Shtabovenko}\ \emph {et~al.}(2023)\citenamefont {Shtabovenko}, \citenamefont {Mertig},\ and\ \citenamefont {Orellana}}]{Shtabovenko:2023idz}%
  \BibitemOpen
  \bibfield  {author} {\bibinfo {author} {\bibfnamefont {V.}~\bibnamefont {Shtabovenko}}, \bibinfo {author} {\bibfnamefont {R.}~\bibnamefont {Mertig}}, \ and\ \bibinfo {author} {\bibfnamefont {F.}~\bibnamefont {Orellana}},\ }\href@noop {} {\  (\bibinfo {year} {2023})},\ \Eprint {http://arxiv.org/abs/2312.14089} {arXiv:2312.14089 [hep-ph]} \BibitemShut {NoStop}%
\bibitem [{\citenamefont {Shtabovenko}\ \emph {et~al.}(2016)\citenamefont {Shtabovenko}, \citenamefont {Mertig},\ and\ \citenamefont {Orellana}}]{Shtabovenko:2016sxi}%
  \BibitemOpen
  \bibfield  {author} {\bibinfo {author} {\bibfnamefont {V.}~\bibnamefont {Shtabovenko}}, \bibinfo {author} {\bibfnamefont {R.}~\bibnamefont {Mertig}}, \ and\ \bibinfo {author} {\bibfnamefont {F.}~\bibnamefont {Orellana}},\ }\href {\doibase 10.1016/j.cpc.2016.06.008} {\bibfield  {journal} {\bibinfo  {journal} {Comput. Phys. Commun.}\ }\textbf {\bibinfo {volume} {207}},\ \bibinfo {pages} {432} (\bibinfo {year} {2016})},\ \Eprint {http://arxiv.org/abs/1601.01167} {arXiv:1601.01167 [hep-ph]} \BibitemShut {NoStop}%
\bibitem [{\citenamefont {Shtabovenko}\ \emph {et~al.}(2020)\citenamefont {Shtabovenko}, \citenamefont {Mertig},\ and\ \citenamefont {Orellana}}]{Shtabovenko:2020gxv}%
  \BibitemOpen
  \bibfield  {author} {\bibinfo {author} {\bibfnamefont {V.}~\bibnamefont {Shtabovenko}}, \bibinfo {author} {\bibfnamefont {R.}~\bibnamefont {Mertig}}, \ and\ \bibinfo {author} {\bibfnamefont {F.}~\bibnamefont {Orellana}},\ }\href {\doibase 10.1016/j.cpc.2020.107478} {\bibfield  {journal} {\bibinfo  {journal} {Comput. Phys. Commun.}\ }\textbf {\bibinfo {volume} {256}},\ \bibinfo {pages} {107478} (\bibinfo {year} {2020})},\ \Eprint {http://arxiv.org/abs/2001.04407} {arXiv:2001.04407 [hep-ph]} \BibitemShut {NoStop}%
\bibitem [{\citenamefont {Mertig}\ \emph {et~al.}(1991)\citenamefont {Mertig}, \citenamefont {Bohm},\ and\ \citenamefont {Denner}}]{Mertig:1990an}%
  \BibitemOpen
  \bibfield  {author} {\bibinfo {author} {\bibfnamefont {R.}~\bibnamefont {Mertig}}, \bibinfo {author} {\bibfnamefont {M.}~\bibnamefont {Bohm}}, \ and\ \bibinfo {author} {\bibfnamefont {A.}~\bibnamefont {Denner}},\ }\href {\doibase 10.1016/0010-4655(91)90130-D} {\bibfield  {journal} {\bibinfo  {journal} {Comput. Phys. Commun.}\ }\textbf {\bibinfo {volume} {64}},\ \bibinfo {pages} {345} (\bibinfo {year} {1991})}\BibitemShut {NoStop}%
\bibitem [{\citenamefont {Shtabovenko}(2017)}]{Shtabovenko:2016whf}%
  \BibitemOpen
  \bibfield  {author} {\bibinfo {author} {\bibfnamefont {V.}~\bibnamefont {Shtabovenko}},\ }\href {\doibase 10.1016/j.cpc.2017.04.014} {\bibfield  {journal} {\bibinfo  {journal} {Comput. Phys. Commun.}\ }\textbf {\bibinfo {volume} {218}},\ \bibinfo {pages} {48} (\bibinfo {year} {2017})},\ \Eprint {http://arxiv.org/abs/1611.06793} {arXiv:1611.06793 [physics.comp-ph]} \BibitemShut {NoStop}%
\end{thebibliography}%

\end{document}